\documentstyle[12pt]{article}
\catcode`\@=11
\def\makepreprititle{\par
  \begingroup
  \def\thefootnote{\fnsymbol{footnote}}
  \def\
@makefnmark{\hbox
  to 0pt{$^{\@thefnmark}$\hss}}
\newpage
  \global\@topnum\z@
  \@makepreprititle \thispagestyle{empty}\@thanks
  \endgroup
  \setcounter{footnote}{0}
  \let\makepreprititle\relax
  \let\@makepreprititle\relax
  \gdef\@thanks{}\gdef\@author{}\gdef\@title{}
  \gdef\@preprintnumber{}\gdef\@preprintdate{}\gdef\subtitle{}
  \let\thanks\relax}
\def\preprintnumber#1{\gdef\@preprintnumber{#1}}
\def\preprintdate#1{\gdef\@preprintdate{#1}}
\def\subtitle#1{\gdef\@subtitle{#1}}
\def\@makepreprititle{\newpage
{\baselineskip 14pt
  \begin{flushright} \@preprintnumber \par
  \@preprintdate \end{flushright} } \par
  \begin{center}
\vskip 1.5em
  {\LARGE \@title \par} \vskip 2.5em
  {\Large \lineskip .5em
  \begin{tabular}[t]{c}\@author
  \end{tabular}\par}
  \vskip 1em {\large \@date} \end{center}
  \par
  \vfil}
\date{\sl Department of Physics, Tohoku University\\Sendai, 980 Japan}
\preprintdate{}
\preprintnumber{}
\subtitle{}
\def\abstract{
\normalsize
\begin{center}
{\bf Abstract\vspace{-.5em}\vspace{0pt}}
\end{center}
\quotation
\addtocounter{page}{-1}
}
\def\endabstract{\if@twocolumn\else\endquotation\fi}
\def\spacing#1{\def\baselinestretch{#1}
\typeout{baselinestretch is modified to \baselinestretch}}
\catcode`\@=12
\spacing{1.4}
\newcommand{\lsim}{\mbox{ \raisebox{-1.0ex}{$\stackrel{\textstyle <}
{\textstyle \sim}$ }}}
\newcommand{\MeV}{\mbox{MeV}}
\newcommand{\GeV}{\mbox{GeV}}
\newcommand{\TeV}{\mbox{TeV}}
\renewcommand{\theequation}{\arabic{section}.\arabic{equation}}
\renewcommand{\thefootnote}{\fnsymbol{footnote}}
\preprintnumber{TU-400}
\preprintdate{July, 1992}
\title{Nucleon Decay\\ in the Minimal Supersymmetric \\
$SU(5)$ Grand Unification}
\author{J.~Hisano, H.~Murayama, and T.~Yanagida}
\begin{document}
\makepreprititle

%\newpage
\begin{abstract}
We make a detailed analysis on the nucleon decay in the minimal supersymmetric
$SU(5)$ grand unified model. We find that a requirement of the unification of
three gauge coupling constants leads to a constraint on a mass $M_{H_C}$ of
color-triplet Higgs multiplet as $2 \times 10^{13}~\GeV \leq M_{H_C} \leq 2
\times 10^{17}~\GeV$, taking both weak- and GUT-scale threshold effects into
account.  Contrary to the results in the previous analyses, the present
experimental limits on the nucleon decay turn out to be consistent with the
SUSY particles lighter than 1~TeV even without a cancellation between matrix
elements contributed from different generations, if one adopts a relatively
large value of $M_{H_C}$ ($\ge2\times 10^{16}~\GeV$).  We also show that the
Yukawa coupling constant of color-triplet Higgs multiplet does not necessarily
blow up below the gravitational scale ($2.4\times10^{18}~\GeV$) even with the
largest possible value of $M_{H_C}$.  We point out that the no-scale model is
still viable, though it is strongly constrained.
\end{abstract}
\newpage

\section{Introduction}
\setcounter{equation}{0}
The hierarchy problem has been the most serious problem in the grand unified
theory (GUT) \cite{GG}. At present, the only feasible solution to this problem
is to introduce the supersymmetry \cite{naturalness}. Furthermore, the
supersymmetric (SUSY) $SU(5)$ model \cite{SUSY-GUT} is now strongly supported
phenomenologically by the $\sin^2 \theta_W$ measurement \cite{revival} made at
the LEP experiments \cite{LEP}.  Once we regard the SUSY-GUT as a serious
candidate of the physics beyond the standard model, a natural question is how
we can test the model. The most striking consequence of the grand unification
is the instability of nucleons. However, the nucleon decay via exchanges of $X$
and $Y$ gauge bosons is strongly suppressed as $\tau^{-1}_{n,p} \propto
M_{GUT}^{-4}$ in the SUSY-GUT because of the large unification scale
\begin{equation}
	M_{GUT} \sim 2 \times 10^{16}~\mbox{GeV}.
\end{equation}
On the other hand, the nucleon decay via exchanges of color-triplet Higgs
multiplet \cite{SYW}, which is suppressed only by $M_{GUT}^{-2}$, may still
allow us to verify the model in the near future.

The main purpose of this paper is to study the implication of the present
experimental limits on the nucleon decay in the minimal SUSY $SU(5)$ GUT
(MSGUT).  Similar analyses have been carried out by Ellis, Nanopoulos, and
Rudaz \cite{ENR}, and later by Arnowitt, Chamseddine, and Nath \cite{AN1,AN2}
rather thoroughly. However, there has been no criterion given on how heavy the
color-triplet Higgs multiplet can be. In this paper, we examine the
experimental limits in the most conservative way, making the color-triplet
Higgs multiplet as heavy as we can, allowed from the renormalization group (RG)
analysis of the gauge coupling constant unification \cite{HMY}. As a
consequence, we find weaker constraints than those given in the previous
analyses. The authors of Refs.~\cite{AN2} have claimed that the data of
nucleon-decay experiments at that time are already stringent enough so that the
SUSY particles below 1~TeV are excluded unless there is a delicate cancellation
between the proton decay matrix elements from second- and third-generation
contributions. On the contrary, we find that the present limits from the
nucleon-decay experiments are still consistent with the SUSY particles below
1~TeV even without such a cancellation. We also study a possible reach of the
superKAMIOKANDE experiment. It will be shown that superKAMIOKANDE, together
with LEP-\uppercase\expandafter{\romannumeral2}, is capable of covering most of
the region with SUSY particles below 1~TeV, and hence it is highly expected to
observe the nucleon decay at superKAMIOKANDE. It will be also stressed that
more precise measurements on the gauge coupling constants, especially on that
of QCD, will give a strong impact on the determination of the color-triplet
Higgs mass $M_{H_C}$.

The paper is organized as follows. A brief review on the MSGUT is presented in
Sect.~2, to summarize our conventions. We critically re-examine the analysis of
Refs.~\cite{AN1,AN2} in Sect.~3. We find that the coefficients of the
dimension-five operators are larger than theirs by a factor of 2. The decay
rates for various modes are presented. As pointed out in Ref.~\cite{AN2}, there
may occur a cancellation between second- and third-generation contributions.
We present the partial lifetimes in terms of unknown parameters $y^{tK}$ or
$y^{t\pi}$ which represent the ratios of the third- to the second-generation
contributions. In Sect.~4, we give an upper bound on the mass of color-triplet
Higgs multiplet, requiring that the gauge coupling constants are unified. Then
the present experimental limits are examined in Sect.~5. There it is explicitly
shown that the present data still allow for SUSY particles below 1~TeV, even
without the cancellation between matrix elements mentioned above. The reach of
the superKAMIOKANDE and the LEP-\uppercase\expandafter{\romannumeral2}
experiments is discussed in Sect.~6.  Sect.~7 is devoted to conclusions and
discussions. An analysis on the dimension-six operators is presented in
Appendix A. Appendix B summarizes discussions on the renormalization effects on
the dimension-five operators. We improve the analysis given in Ref.~\cite{ENR},
but the difference turns out to be small.  The chiral Lagrangian technique
adopted to calculate the nucleon-decay matrix elements is described in Appendix
C.

\section{Minimal SUSY $SU(5)$ GUT}
\setcounter{equation}{0}
In this section, we review the minimal SUSY $SU(5)$ GUT (MSGUT) \cite{DGS},
summarizing our conventions. We also clarify the origin of new CP-violating
phases in Yukawa coupling constants of color-triplet Higgs multiplet to matter
multiplets.

There are quite a few multiplets in the MSGUT. An adjoint Higgs multiplet
$\Sigma(\mbox{\bf 24})$ breaks the $SU(5)$ GUT group down to $SU(3)_C
\times SU(2)_L \times U(1)_Y$, and a pair of quintets $H(\mbox{\bf 5})$ and
$\overline{H} (\mbox{\bf 5}^\ast)$ contain doublet Higgs multiplets
$H_f,\,\overline{H}_f$ in the minimal SUSY standard model as well as their
color-triplet partners $H_C,\,\overline{H}_C$. The superpotential of this model
is
\begin{eqnarray}
W &=& \frac{f}{3} {\rm Tr} \Sigma^3
	+ \frac{1}{2} fV {\rm Tr} \Sigma^2
	+ \lambda \overline{H}_\alpha
		(\Sigma^\alpha_\beta +3V \delta^\alpha_\beta) H^\beta
			\nonumber \\
 & &	+ \frac{h^{ij}}{4} \varepsilon_{\alpha\beta\gamma\delta\epsilon}
		 \psi_i^{\alpha\beta} \psi_j^{\gamma\delta} H^\epsilon
	+ \sqrt{2} f^{ij} \psi_i^{\alpha\beta} \phi_{j\alpha}
		\overline{H}_\beta,
\label{superpotential}
\end{eqnarray}
where the Latin indices $i,j = 1, 2, 3$ refer to families, and the Greek ones
$\alpha, \beta, \gamma \cdots$ represent the $SU(5)$ indices.  The chiral
superfields $\psi$({\bf 10}), $\phi$({\bf 5}$^\ast$) are matter multiplets.
Contents of the Higgs multiplets are
\begin{eqnarray}
	\Sigma &=& \Sigma^a T^a \nonumber\\
               &=& \left( \begin{array}{cc}
                     \Sigma_8 & \Sigma_{(3,2)} \\
	             \Sigma_{(3^*,2)} & \Sigma_3 \\
                    \end{array} \right)
		+ \frac{1}{2\sqrt{15}} \left( \begin{array}{cc}
			2 & 0 \\ 0 & -3
					      \end{array}
					\right) \Sigma_{24}, \\
	{}^t H &=& (H_C,\,H_C,\,H_C,\,H_f^+,\,H_f^0), \\
	{}^t \overline{H} &=& (\overline{H}_C,\,\overline{H}_C,\,
		\overline{H}_C,\,\overline{H}_f^-,\,-\overline{H}_f^0),
\end{eqnarray}
and those of the matter multiplets are
\begin{eqnarray}
	\psi &=& \frac{1}{\sqrt{2}}
		\left( \begin{array}{ccccc}
			0 & u^c & -u^c & u & d \\
			-u^c & 0 & u^c & u & d \\
			u^c & -u^c & 0 & u & d \\
			-u & -u & -u & 0 & e^c \\
			-d & -d & -d & -e^c & 0
		\end{array} \right), \nonumber \\
	{}^t \phi &=& (d^c,\,d^c,\,d^c,\,e,\,-\nu).
\end{eqnarray}
where all the matter multiplets are written in terms of the chiral
(left-handed) superfields. The chiral superfields $u$ and $d$ contain
left-handed up-type and down-type quarks, $u^c$ and $d^c$ the charge
conjugations of right-handed up-type and down-type quarks, $e$ and $\nu$
left-handed charged leptons and neutrinos, and $e^c$ the charge conjugations of
right handed charged-leptons.  In the following, ${}^t Q \equiv (u,d)$ and
${}^t L\equiv (\nu,e)$ will denote chiral superfields of weak-doublet quarks
and leptons, respectively.

The $SU(5)$ GUT symmetry is broken by a vacuum expectation value of the
$\Sigma$ field,
\begin{equation}
\langle \Sigma \rangle = V \left( \begin{array}{ccccc}
		2&&&&\\
		&2&&&\\
		&&2&&\\
		&&&-3&\\
		&&&&-3	\end{array} \right),
\end{equation}
giving masses to $X$ and $Y$ gauge bosons
\begin{equation}
	M_V \equiv M_X = M_Y = 5 \sqrt2 g_5 V,
\end{equation}
where $g_5$ is the unified $SU(5)$ gauge coupling constant. The invariant mass
parameter of $H$ and $\overline{H}$ is fine-tuned to
realize masslessness of $H_f$ and $\overline{H}_f$, while it keeps
their color-triplet partners, $H_C$ and $\overline{H}_C$, superheavy as
\begin{equation}
		M_{H_C} = M_{\overline{H}_C} = 5 \lambda V.
			\label{MHCmass}
\end{equation}
The components $\Sigma_8$ and $\Sigma_3$ acquire the same mass
\begin{equation}
       M_{\Sigma} \equiv M_{\Sigma_8} = M_{\Sigma_3} = \frac52 f V,
                        \label{24mass}
\end{equation}
while the (physical) components $\Sigma_{(3^*,2)}$ and $\Sigma_{(3,2)}$ form
superheavy vector multiplets of mass $M_V$ together with the gauge multiplets.
The mass of the singlet component $\Sigma_{24}$ is $(1/2) fV$.

To analyze the dimension-five operators, we have to examine the Yukawa
couplings of $H$ and $\overline{H}$ to matter multiplets. An important question
is how many independent parameters we have in the Yukawa couplings
\cite{EGN}. The Yukawa coupling constants $h^{ij}$ and $f^{ij}$
in Eq.~(\ref{superpotential}) form a parameter space {\bf C}$^6 \times$ {\bf
C}$^9$, since $h^{ij}$ is a symmetric matrix. The freedom of field
re-definition is $U(3) \times U(3)$, corresponding to the choice of the basis
of $\psi_i$ and $\phi_i$. Thus the physical degrees of freedom of the Yukawa
coupling constants is $(6 + 9)\times 2 - 9 \times 2 = 12 = 3 + 3 + 4 + 2$.
First two $3$'s stand for the eigenvalues for up- and down-type mass matrices,
$4$ for the Kobayashi-Maskawa matrix elements, and $2$ for the additional
phase degrees of freedom.  We will parameterize the coupling matrices $h^{ij}$
and $f^{ij}$ as
\begin{eqnarray}
	h^{ij} &=& h^i e^{i \varphi_i} \delta^{ij}, \\
	f^{ij} &=& V_{ij}^\ast f^j,
\end{eqnarray}
with $V_{ij}$ being the Kobayashi-Maskawa matrix. Only two of the phases $e^{i
\varphi_i}$ are independent, and we can take
\begin{equation}
	\varphi_u + \varphi_c + \varphi_t = 0.
\end{equation}
In this parameterization, the corresponding bases of the matter multiplets are
\begin{eqnarray}
	\psi_i &\ni& {}^t Q_i \equiv (u_i,\,d'_i) = (u_i,\,V_{ij} d_j),\\
	\psi_i &\ni& e^{-i \varphi_i} u_i^c, \\
	\psi_i &\ni& V_{ij} e_j^c, \\
	\phi_i &\ni& d_i^c, \\
	\phi_i &\ni& {}^t L_i \equiv (\nu_i,e_i),
\end{eqnarray}
in terms of the mass eigenstates
$u_i,\,d_i,\,u_i^c,\,d_i^c,\,\nu_i,\,e_i,\,e_i^c$. Then the Yukawa couplings of
Higgs to matter multiplets are given by
\begin{eqnarray}
W_Y &=&  h^i Q_i u_i^c H_f
      +  V_{ij}^\ast f^j Q_i d_j^c \overline{H}_f+ f_i e_i^c L_i \overline{H}_f
      \\
   && +  \frac12 h^i e^{i \varphi_i} Q_i Q_i H_C
      +  V_{ij}^\ast f^j Q_i L_j \overline{H}_C
     \nonumber\\
   && +  h^i  V_{ij} u_i^c e_j^c H_C
      +  e^{ - i \varphi_i} V_{ij}^\ast f^j u_i^c d_j^c \overline{H}_C.
     \nonumber
\end{eqnarray}
It should be clear from the above expression that the phases $e^{i \varphi_i}$
would be completely irrelevant if $H_C$ and $\overline{H}_C$ were absent. The
phases appearing in the Yukawa couplings of $H_C$ and $\overline{H}_C$ cannot
be absorbed by the field re-definition without affecting the couplings of $H_f$
and $\overline{H}_f$. As we will see later, these phases are important in the
nucleon-decay amplitudes induced by the $H_C$ and $\overline{H}_C$ exchanges.
However, they are perfectly irrelevant to the nucleon decay caused by the $X$
and $Y$ gauge-boson exchanges.

\section{Dimension-Five Operators and Decay Rates}
\setcounter{equation}{0}
In this section, we re-examine the previous analyses by Ellis, Nanopoulos, and
Rudaz \cite{ENR}, and by Arnowitt, Chamseddine, and Nath \cite{AN1,AN2}.
Several corrections to the formula and numerical factors are made, and as a
consequence nucleon-decay amplitude turns out to be smaller than their result
by a factor of 2.

In the SUSY-GUT there are several baryon-number violating operators since it
has many scalar bosons with color quantum numbers. Dimension-six operators
induced by the $X$ and $Y$ gauge-boson exchanges are suppressed by
$1/M_{GUT}^2$. These operators cause unacceptably large nucleon-decay rates in
the minimal non-SUSY $SU(5)$ GUT \cite{mar}, but there is no problem in the
MSGUT since $M_{GUT}$ is much larger than in the non-SUSY case. In fact, we
have analyzed the nucleon decays caused by the dimension-six operators, and
found that they are always suppressed compared to those caused by the
dimension-five operators. A brief discussion on the dimension-six operators is
presented in Appendix A. The dimension-five operators are much more dangerous.
These operators are generated by $H_C$ and $\overline{H}_C$ exchanges in the
MSGUT, and is suppressed only by $1/M_{GUT}$ \cite{SYW}. The nucleon-decay
amplitudes are obtained by dressing these operators by SUSY particle exchanges
to convert scalar bosons to light fermions. Therefore, the nucleon-decay rates
are sensitive to SUSY particle masses as well as the mass of $H_C$ and
$\overline{H}_C$. It has been also noted that there may be baryon-number
violating dimension-four operators \cite{DG}. However, they can be forbidden by
imposing the $R$-parity invariance, and we will not consider them in this
paper.

Let us now discuss the dimension-five operators that cause the nucleon decay. A
supergraph is presented in Fig.~1. The operators can be written explicitly as
\begin{equation}
W_5 = \frac{1}{2M_{H_C}} h^{i} e^{i \varphi_i} V_{kl}^\ast f^l
	(Q_i Q_i) (Q_k L_l)
    + \frac{1}{M_{H_C}} h^i V_{ij} e^{-i \varphi_k} V_{kl}^\ast f^l
        (u^c_i e^c_j) (u^c_k d^c_l),
		\label{dimen5}
\end{equation}
where the contraction of the indices are understood as
\begin{eqnarray}
	(Q_i Q_i) (Q_k L_l) &=& \varepsilon_{\alpha \beta \gamma}
		(u_i^\alpha d_i^{\prime\beta} -
			d_i^{\prime\alpha} u_i^\beta )
		(u_k^\gamma e_l - d_k^{\prime\gamma} \nu_l), \\
	(u^c_i e^c_j) (u^c_k d^c_l) &=& \varepsilon^{\alpha \beta \gamma}
		u^c_{i\alpha} e^c_j u^c_{k\beta} d^c_{l\gamma}.
\end{eqnarray}
with $\alpha$, $\beta$, $\gamma$ being color indices. Note that the total
anti-symmetry in the color index requires that the operators are flavor
non-diagonal $(i \neq k)$. Therefore dominant decay modes in the MSGUT
generally involve strangeness, like $n$, $p \rightarrow K \bar{\nu}$
\cite{DRW,ENR}.

The dimension-five operators will be converted to four-fermi operators at the
SUSY breaking scale, by exchanges of gauginos or doublet Higgsino. However, the
important contributions come only from the charged-wino dressing of the $(Q_i
Q_i) (Q_k L_l)$ operators, and we will concentrate to this case.

The exchanges of gluino, neutral gaugino and neutral Higgsino are small in
general \cite{AN2}, since they are flavor diagonal and hence suppressed by the
Yukawa coupling constants of first and second generations appearing in the
dimension-five operators. Though the gluino exchanges have stronger gauge
coupling $\alpha_3$ than the wino exchanges, it will vanish completely in the
limit where all the squark masses are degenerate \cite{DRW}.  Since the high
degeneracy is required to suppress the unwanted flavor changing neutral
current,\footnote{A high degeneracy is required to suppress flavor-changing
neutral currents, especially between first and second generations.  See
Ref.~\cite{FCNC}.} the gluino exchanges turn out to be always small. The
charged-Higgsino exchanges are also suppressed due to their small Yukawa
coupling constants to the first or second generations.\footnote{There are
contributions from the third generation to the charged-Higgsino exchange
amplitudes. However, the bottom-quark Yukawa coupling constant $f^b$
is smaller than
the $SU(2)$ gauge coupling constant $g_2$ unless $\tan\beta_H$ in
Eq.~(\ref{tanb}) is extremely large. Therefore, the charged-Higgsino exchanges
are most likely smaller than the charged-wino exchanges.} Thus the charged-wino
exchanges give dominant contributions. Note that charged-wino dressing is
impossible for the second operators $(u^c_i e^c_j) (u^c_k d^c_l)$, since all
the fields involved are right-handed fields. Though the left-right mixing due
to the $A$-terms and superpotential $|\partial W/\partial H|^2$ induces the
wino dressing to the operators $(u^c_i e^c_j) (u^c_k d^c_l)$, their
contribution is suppressed similarly to the charged-Higgsino exchanges since
the mixing is proportional to the Yukawa coupling constants. In the following
we refer to the charged-wino simply as wino.

By dressing the dimension-five operators with the wino exchanges, we will get
four-fermi operators. The results depend on the masses of charginos, squarks,
and sleptons in the loops. There is a mixing between wino and charged Higgsino,
with the mass matrix
\begin{equation}
	M_{\rm chargino} =
     \left( \begin{array}{cc}
            m_{\tilde{w}} & \sqrt{2} m_W \cos \beta_H \\
	    \sqrt{2} m_W \sin \beta_H & \mu \\
             \end{array} \right).
\label{chargino}
\end{equation}
Here, $m_{\tilde{w}}$ is a pure wino mass, $\mu$ a pure Higgsino mass, and
$\beta_H$ a vacuum angle of doublet Higgs scalars, defined by
\begin{equation}
\tan\beta_H = \frac{\langle H_f^0 \rangle}{\langle \overline{H}_f^0 \rangle}.
\label{tanb}
\end{equation}
The interaction of squarks and sleptons with wino
is fixed by
\begin{equation}
	{\cal L} = g_2 (\tilde{u}_L^\ast \tilde{w}^+ d_L
			+ \tilde{d}_L^\ast \tilde{w}^- u_L
                        + \tilde{\nu}_L^\ast \tilde{w}^+ e_L
			+ \tilde{e}_L^\ast \tilde{w}^- \nu_L) + \mbox{h.c.},
\end{equation}
giving the triangle diagram factor \cite{AN2}
\begin{equation}
       \frac{\alpha_2}{2\pi} f(u,\,d) \equiv
        g_2^2\int \frac{d^4 k}{i(2\pi)^4}
       \left( \frac{1}{m_{\tilde{u}}^2 - k^2} \right)
       \left( \frac{1}{m_{\tilde{d}}^2 - k^2} \right)
       \left( \frac{1}{M_{\rm chargino} - {\not\!k} } \right)_{11}.
\label{tri}
\end{equation}
We have taken only the $(1,1)$ component of the chargino propagator, since the
nucleon-decay amplitudes are dominated by the pure wino component in the
chargino states.  Though the integral Eq.~(\ref{tri}) depends on the mass
eigenvalues and the mixing angles, we have found that it is well approximated
by the pure wino exchange, and then it can be given by
\begin{equation}
         \frac{\alpha_2}{2\pi} f(u,\,d) \equiv
          \frac{\alpha_2}{2\pi} m_{\tilde{w}}
		\frac{1}{m_{\tilde{u}}^2 - m_{\tilde{d}}^2}
		\left( \frac{m_{\tilde{u}}^2}{m_{\tilde{u}}^2 -m_{\tilde{w}}^2}
			\ln \frac{m_{\tilde{u}}^2}{m_{\tilde{w}}^2}
		- \frac{m_{\tilde{d}}^2}{m_{\tilde{d}}^2 - m_{\tilde{w}}^2}
			\ln \frac{m_{\tilde{d}}^2}{m_{\tilde{w}}^2} \right).
\label{fff}
\end{equation}
This approximation can be easily justified if $m_{\tilde{w}} \gg m_W$, since
then the off-diagonal elements in the mass matrix Eq.~(\ref{chargino}) can be
neglected. On the other hand, if $m_{\tilde{w}} \sim m_W$, the off-diagonal
elements cannot be neglected in general. However, if $m_{\tilde{Q}},
m_{\tilde{L}} \gg m_{\tilde{w}}$, the triangle diagram factor $f$ in
Eq.~(\ref{tri}) is simply
given by
\begin{equation}
f \simeq \frac{(M_{\rm chargino})_{11}}{m_{\tilde{Q}}^2}
	= \frac{m_{\tilde{w}}}{m_{\tilde{Q}}^2},
\label{squarkheavy}
\end{equation}
and hence the above approximation is again justified. The region $m_{\tilde{w}}
\sim m_{\tilde{Q}} \sim m_{\tilde{L}} \sim \mu \sim m_{W}$ requires an exact
treatment of the mixing, but this region turns out to be already excluded, and
is irrelevant to our analyses.

Notice that the nucleon decay rates depend on the SUSY particle masses only
through the function $f$. It is useful to see the dependence of the function
$f$ on $m_{\tilde{w}}$, $m_{\tilde{Q}}$, and $m_{\tilde{L}}$. In the limit
$m_{\tilde{w}} \ll m_{\tilde{Q}}\sim m_{\tilde{L}}$, $f$ behaves as in
Eq.~(\ref{squarkheavy}). In the other limit $m_{\tilde{w}} \gg
m_{\tilde{Q}}\sim m_{\tilde{L}}$, it behaves as
\begin{equation}
f \simeq \frac{1}{m_{\tilde{w}}} \ln\frac{m_{\tilde{w}}^2}{m_{\tilde{Q}}^2}.
\end{equation}
These behaviors will be used to put bounds on these masses in section 5.

The resulting four-fermi operators can be written down explicitly as
\begin{eqnarray}
\lefteqn{
	{\cal L} = \frac{1}{M_{H_C}} \frac{\alpha_2}{2\pi}
		h^i e^{i \varphi_i} V_{jk}^\ast f^k
		\varepsilon_{\alpha\beta\gamma}
			} \nonumber \\
& & \times \left[ (u_i^\alpha d_i^{\prime\beta}) (d_j^{\prime\gamma} \nu_k)
				(f(u_j,\,e_k) + f(u_i,\,d'_i))
	+ (d_i^{\prime\alpha} u_i^\beta) (u_j^\gamma e_k)
				(f(u_i,\,d_i) + f(d'_j,\,\nu_k))
		\right. \nonumber \\
& & \left.
	+ (d_i^{\prime\alpha} \nu_k) (d_i^{\prime\beta} u_j^\gamma)
				(f(u_i,\,e_k) + f(u_i,\,d'_j))
	+ (u_i^\alpha d_j^{\prime\beta}) (u_i^\gamma e_k)
				(f(d'_i,\,u_j) + f(d'_i,\,\nu_k)) \right],
			\nonumber \\
\label{fourf}
\end{eqnarray}
where the contraction of spinor indices are taken in each brackets $()$. Here
we have assumed that the mixing between squarks is negligible. This is true in
most of the supergravity models (for example, see Ref.\cite{nilles})
which ensure the absence of flavor-changing neutral current. Notice that
Eq.~(\ref{fourf}) is larger than that given in Refs.~\cite{AN1,AN2} by a factor
of 2. We suspect that this difference arises from an inconsistency of the
normalization of their Yukawa coupling constants (See Eqs.~(1.5), (1.6) in
Ref.~\cite{AN1}).

We have to include three kinds of renormalization effects to perform
quantitative analyses. First, the Yukawa coupling constants appearing in the
Eq.~(\ref{dimen5}) are those evaluated at the GUT-scale, and we have to
calculate their magnitudes using the low-energy quark masses. Second, the
dimension-five operators receive anomalous dimensions due to the wave-function
renormalizations of the external lines. Third, the four-fermi operators
obtained after the gaugino-dressing will be further renormalized from the SUSY
breaking scale down to 1~GeV. These three effects are first discussed by Ellis,
Nanopoulos, and Rudaz \cite{ENR}. However, they dropped the $SU(2)$ gauge
interactions in estimating the first renormalization effects, which led to an
overestimation by 50\%. They also neglected the contributions from the
top-quark Yukawa coupling. Since the top-quark mass is heavier than $90~\GeV$,
these contributions, which appear in the wave-function renormalization of Higgs
doublets, enhance the dimension-five operators by $\sim 30\%$. However,
corrections on these two points give roughly the same result as theirs. The
details are explained in the Appendix B.

After taking the renormalization effects into account, the nucleon-decay
operators at 1~GeV are given by
\begin{eqnarray}
\lefteqn{
	{\cal L} = \frac{2\alpha_2^2}{M_{H_C}}
	\frac{\overline{m}_{u_i} \overline{m}_{d_k}
		e^{i \varphi_i} V_{jk}^\ast}{m_W^2 \sin 2\beta_H}
		A_S(i,j,k) A_L } \nonumber \\
& &	\times \varepsilon_{\alpha\beta\gamma}
	\left[ (u_i^\alpha d_i^{\prime\beta}) (d_j^{\prime\gamma} \nu_k)
				(f(u_j,\,e_k) + f(u_i,\,d'_i))
	+ (d_i^{\prime\alpha} u_i^\beta) (u_j^\gamma e_k)
				(f(u_i,\,d_i) + f(d'_j,\,\nu_k))
			\right. \nonumber \\
& &	\left.
	+ (d_i^{\prime\alpha} \nu_k) (d_i^{\prime\beta} u_j^\gamma)
				(f(u_i,\,e_k) + f(u_i,\,d'_j))
	+ (u_i^\alpha d_j^{\prime\beta}) (u_i^\gamma e_k)
				(f(d'_i,\,u_j) + f(d'_i,\,\nu_k)) \right],
		\nonumber \\	\label{dim6}
\end{eqnarray}
where $A_S(i,j,k)$ represents the short-range renormalization effect between
GUT- and SUSY breaking scales depending on the flavor $i,j,k$, and $A_L$ the
long-range renormalization effect between SUSY scale and 1~GeV. Here, the quark
masses $\overline{m}_{u_i}$ and $ \overline{m}_{d_k}$ are defined at $1~\GeV$
in the $\overline{MS}$ scheme \cite{ENR}.

We first show the prediction of nucleon-decay rates for the dominant modes, $n,
p \rightarrow K \bar{\nu}_\mu$.  The main contributions in the four-fermi
operators Eq.~(\ref{dim6}) come from the terms $(i=c,\,j=u,\,k=s)$
(proportional to $\overline{m}_{c} \overline{m}_{s}$), and $(i=t,\,j=u,\,k=s)$
(proportional to $\overline{m}_{t} \overline{m}_{s}$). The relevant terms are
given by
\begin{eqnarray}
{\cal L} &=& \frac{2\alpha_2^2}{M_{H_C}}
	\frac{\overline{m}_s V_{us}^\ast A_L}{m_W^2 \sin 2\beta_H}
	\varepsilon_{\alpha\beta\gamma}
	\left((d^\alpha u^\beta) (s^\gamma \nu_\mu)
			+ (s^\alpha u^\beta)(d^\gamma \nu_\mu) \right)
		\nonumber\\
& & \times \left[ A_S(c,u,s) \overline{m}_c
		e^{i\varphi_c} V_{cs} V_{cd} (f(c,\,\mu) + f(c,\,d'))
    \right. \nonumber\\
& & \left.     	+ A_S(t,u,s) \overline{m}_t
		e^{i\varphi_t} V_{ts} V_{td} (f(t,\,\mu) + f(t,\,d'))
    \right].
\end{eqnarray}
We have neglected the terms propotional to $\overline{m}_u$.  Though the terms
coming from the $\tilde{c}$-exchange can be computed precisely, the
contribution of $\tilde{t}$-exchange is ambiguous due to the unknown
Kobayashi-Maskawa matrix elements for top quark \cite{PDG}. In fact, the
ratio of the $\tilde{t}$-contribution relative to the $\tilde{c}$-one
\cite{AN1},
\begin{equation}
y^{tK} \equiv
\frac{\overline{m}_t e^{i\varphi_t} V_{ts} V_{td} A_S(t,u,s)
	(f(t,\,\mu) + f(t,\,d'))}
	{\overline{m}_c e^{i\varphi_c} V_{cs} V_{cd} A_S(c,u,s)
	(f(c,\,\mu) + f(c,\,d'))}
\label{ytk}
\end{equation}
ranges between
\begin{equation}
		0.096 < |y^{tK}| < 1.3,
\end{equation}
if we take the triangle diagram factors $f$ to be common, and the top-quark
mass $m_t$ to be $100~\GeV$.\footnote{Note that $\overline{m}_t$ in
Eq.~(\ref{ytk}) is defined at the renormalization point $1~\GeV$. For example,
$\overline{m}_t=270~\GeV$ for $m_t=100~\GeV$.}

Note that $e^{i\varphi_t}$ and $e^{i\varphi_c}$ in Eq.~(\ref{ytk}) are
independent of each other. We cannot measure these phases from the present-day
experiments, since they are irrelevant to any obsevables as far as $H_C$ and
$\overline{H}_C$ are decoupled. Thus we are completely ignorant whether
$y^{tK}$ is {\it constructive}\/ or {\it destructive}\/ to the
$\tilde{c}$-exchange amplitude. We should regard this {\it complex} parameter
free in the present analyses.

If the modes $n,p \rightarrow K \bar{\nu}_{\mu}$ have a cancellation between
$\tilde{c}$- and $\tilde{t}$-exchange amplitudes, we have to study the other
possible decay modes. Next-leading modes are $n,p \rightarrow \pi
\bar{\nu}_{\mu}$, suppressed by the Cabbibo angle $\sin^2\theta_C$ compared to
the $K \bar{\nu}_{\mu}$ modes.  There are, similarly to the $K \bar{\nu}_{\mu}$
modes, contributions from $\tilde{c}$- and $\tilde{t}$-exchange to these modes,
and their ratio
\begin{equation}
y^{t\pi} \equiv
\frac{\overline{m}_t e^{i\varphi_t} V_{td}^2 A_S(t,u,s)
	(f(t,\,\mu) + f(t,\,d'))}
{\overline{m}_c e^{i\varphi_c} V_{cd}^2 A_S(c,u,s)
	(f(c,\,\mu) + f(c,\,d'))},
\label{ytpi}
\end{equation}
ranges as
\begin{equation}
               0.041< |y^{t\pi}| < 1.7,
\end{equation}
with common $f$ and $m_t = 100~\GeV$. Though $y^{tK}$ and $y^{t\pi}$ are
correlated, we also regard $y^{t\pi}$ as an independent parameter because
of too large uncertainties in the Kobayashi-Maskawa matrix elements
\begin{eqnarray}
\left|\frac{y^{t\pi}}{y^{tK}}\right|&=&
	\left|\frac{V_{td} V_{cs}}{V_{ts} V_{cd}}\right| \\
                       &=& 0.22 -2.94. \nonumber
\end{eqnarray}

We find that a perfect ``double'' cancellation in $K\bar{\nu}$ and
$\pi\bar{\nu}$ modes is not possible, since $y^{tK}$ and $y^{t\pi}$ have a
non-vanishing relative phase coming from the CP-violating one in the
Kobayashi-Maskawa matrix.  Thus, we will not consider the ``double''
cancellation further in this paper, and take $|1 + y^{t\pi}| = 1$ throughout. A
``double'' cancellation is logically possible only if the CP-violation is
dominated by a non-KM mechanism. Even in the presence of such a ``double''
cancellation, there are decay modes which do not have such ambiguities ({\it
i.e.}, $p \rightarrow K^0 \mu^+, \eta^0 \mu^+, \pi^0 \mu^+$ and $n \rightarrow
\pi^- \mu^+$). The operators containing charged leptons have only a
contribution from the up-quark Yukawa coupling constant. Therefore, the decay
rates do not suffer from the ambiguity of a possible cancellation. However, the
decay rates are very small since the up-quark Yukawa coupling constant is tiny.

To obtain matrix elements at the hadronic level from the operators written in
terms of the quarks fields, we adopt a chiral Lagrangian technique
\cite{CWH,cd}.  Details on this method is shown in appendix C. We present the
results on the partial lifetimes of nucleons in Tables~1--3, in terms of the
parameters
\begin{center}
	$\beta, M_{H_C}, A_S, \beta_H, y^{tK}, y^{t\pi},$
\end{center}
and the triangle factors $f$'s. A parameter $\beta$ is the hadron matrix
element parameter used in the chiral Lagrangian technique
\cite{BEHS,GKSM}
\begin{equation}
\beta u_L(\vec{k}) \equiv \epsilon_{\alpha \beta \gamma}
	\langle 0|
		(d_L^{\alpha} u_L^{\beta}) u_L^{\gamma}|p,\vec{k}\rangle,
\label{beta}
\end{equation}
which ranges as
\begin{equation}
	\beta = \mbox{(0.003 -- 0.03)~GeV}^3,
\end{equation}
depending on the methods of the theoretical estimation. Due to the uncertainty
of an order of magnitude, the predictions of nucleon partial lifetimes receive
an ambiguity of two orders of magnitude. A more precise determination of
$\beta$ is strongly desired.

Table~1 summarizes predictions on the nucleon partial lifetimes for the
dominant modes $n,p \rightarrow K \bar{\nu}$.  One sees that the partial
lifetimes of neutron are a little shorter than ones of proton. This is because
the former has a larger chiral Lagrangian factor.  Table~2 presents
next-dominant modes $n,p \rightarrow \pi \bar{\nu}$. These modes are suppressed
by the Cabbibo-angle $\sin^2\theta_C$. Ratio of the decay rates into
$\bar{\nu}_\mu$ and $\bar{\nu}_e$ is simply the squared ratio of strange- and
down-quark masses. The decay rates into charged lepton $\mu^+$ are listed in
Table~3. These decay rates do not suffer from the ambiguity of a possible
cancellation. However, the partial lifetimes are very long because of the small
Yukawa coupling constant of up quark.

The dimension-five operators given by Eq.~(\ref{dim6}) are larger than the
expressions given in Refs.~\cite{AN1,AN2}. However, our final results shown in
Tables~1--3 are smaller than the conclusion in Ref.~\cite{AN2} by a factor of 2
which may originate in an inconsistency between their analytic formula
Eq.~(2.3) and its numerical evaluations Eq.~(2.7) in Ref.~\cite{AN2}.

So far we have not discussed the decay rates of the modes containing
$\bar{\nu}_{\tau}$.  This is because we cannot make definite predictions for
these decay modes since $V_{ub}^*$ has a large ambiguity. While the decay rates
of $\bar{\nu}_{\tau}$ modes can be as large as ones of $\bar{\nu}_{\mu}$ mode
if we take the largest possible $V_{ub}^*$ value, they can be also negligible
with the smallest possible $V_{ub}^*$ value. In any case the decay rates can be
only comparable to those into the $\bar{\nu}_\mu$, and hence the total decay
rate into $\bar{\nu}$ is raised at most by a factor of two. This gives only a
factor of $\sqrt{2}$ stronger constraint on the squark masses, and we will not
include the $\bar{\nu}_{\tau}$ modes, hereafter. Once we know $V_{ub}^\ast$
more precisely, it is easy to incorporate the $\bar{\nu}_{\tau}$ modes into the
present analyses.

Finally, we note that dimension-five operators depend sensitively on quark
masses.  We use the central values of current quark masses at $1~\GeV$ which
are estimated from the chiral perturbation theory and QCD sum rule in
Ref.~\cite{GL}. Quark masses of the first and second generations have large
ambiguities. Especially, the strange-quark mass has a large error-bar ({\it
i.e.}, $\overline{m}_s(1~\GeV) = 175 \pm 55 ~\MeV$, given in the
$\overline{MS}$ scheme). In our analyses we use $\overline{m}_s(1~\GeV) =
175~\MeV$.

\section{Constraints on the GUT-scale Mass Spectrum}
\setcounter{equation}{0}
Since the nucleon-decay rates due to the dimension-five operators are
proportional to the inverse square of the color-triplet Higgs multiplet mass
$M_{H_C}^{-2}$, it is very important to determine it by some means. In the
previous analyses \cite{ENR,AN1,AN2}, the authors have chosen $M_{H_C}=(1-2)
\times 10^{16}~\GeV$ ad hoc. However, we have shown recently \cite{HMY}, that
one can obtain limits on the GUT-scale mass spectrum in the MSGUT, just by
requiring the unification of three gauge coupling constants. In particular, we
have derived the upper bound on $M_{H_C}$ without any theoretical prejudice.  A
theoretical requirement that the Yukawa couplings remain perturbative below the
gravitational scale $M_P/\sqrt{8\pi}$ ($2.4\times10^{18}~\GeV$), poses further
constraint on the GUT-scale mass spectrum.

We first discuss the renormalization-group (RG) evolution of three gauge
coupling constants. It was shown in Refs.~\cite{EJ,AKT} that the simple
step-function approximation is accurate for supersymmetric theories, justified
in the ``supersymmetric regularization'' $\overline{DR}$ scheme \cite{siegel}.
To illustrate how the GUT-scale spectrum receives constraints, we first discuss
the one-loop RG equations. After that we include the two-loop corrections.

The running of three gauge coupling constants in the MSGUT can be obtained
easily at the one-loop level as
\begin{eqnarray}
\alpha_3^{-1} (m_Z) &=& \alpha_{5}^{-1} (\Lambda)
	+ \frac{1}{2\pi} \left\{
		\left( -2 - \frac{2}{3} N_g \right) \ln \frac{m_{SUSY}}{m_Z}
		+ (-9 + 2 N_g) \ln \frac{\Lambda}{m_Z}
			\right. \nonumber \\
& & \phantom{\alpha_{5}^{-1} (\Lambda) + \frac{1}{2\pi} \{ } \left.
		-4 \ln \frac{\Lambda}{M_V} + 3 \ln \frac{\Lambda}{M_\Sigma}
		+ \ln \frac{\Lambda}{M_{H_C}} \right\}, \\
            \alpha_2^{-1} (m_Z) &=& \alpha_{5}^{-1} (\Lambda)
	+ \frac{1}{2\pi} \left\{
		\left( -\frac{4}{3} - \frac{2}{3} N_g - \frac{5}{6} \right)
			 \ln \frac{m_{SUSY}}{m_Z}
		+ (-6 + 2 N_g + 1) \ln \frac{\Lambda}{m_Z}
			\right. \nonumber \\
& & \phantom{\alpha_{5}^{-2} (\Lambda) + \frac{1}{2\pi} \{ } \left.
		-6 \ln \frac{\Lambda}{M_V} + 2 \ln \frac{\Lambda}{M_\Sigma}
		\right\}, \\
\alpha_1^{-1} (m_Z) &=& \alpha_{5}^{-1} (\Lambda)
	+ \frac{1}{2\pi} \left\{
		\left( -\frac{2}{3} N_g - \frac{1}{2} \right)
			 \ln \frac{m_{SUSY}}{m_Z}
		+ \left(2 N_g + \frac{3}{5} \right)
			\ln \frac{\Lambda}{m_Z}
			\right. \nonumber \\
& & \phantom{\alpha_{5}^{-3} (\Lambda) + \frac{1}{2\pi} \{ } \left.
		-10 \ln \frac{\Lambda}{M_V}
		+ \frac{2}{5} \ln \frac{\Lambda}{M_{H_C}} \right\}.
\end{eqnarray}
Here, the scale $\Lambda$ is supposed to be larger than any of the GUT-scale
masses.  The number of generations $N_g$ is three, and we have assumed a common
mass $m_{SUSY}$ for all the SUSY particles and for the scalar component of one
of the Higgs doublets. A mass of the other doublet Higgs boson is taken at
$m_Z$. By eliminating $\alpha^{-1}_{5}$ from the above equations, we obtain
simple relations:
\begin{eqnarray}
(3 \alpha_2^{-1} - 2 \alpha_3^{-1} - \alpha_1^{-1}) (m_Z)
	&=& \frac{1}{2\pi} \left\{
		\frac{12}{5} \, \ln \frac{M_{H_C}}{m_Z}
		- 2 \, \ln \frac{m_{SUSY}}{m_Z} \right\},
			\label{MHC}
			\\
(5 \alpha_1^{-1} - 3 \alpha_2^{-1} - 2 \alpha_3^{-1}) (m_Z)
	&=& \frac{1}{2\pi} \left\{
		12 \, \ln \frac{M_V^2 M_\Sigma}{m_Z^3}
		+ 8 \ln \frac{m_{SUSY}}{m_Z} \right\}.
		\label{MGUT}
\end{eqnarray}
The Eqs.~(\ref{MHC}), (\ref{MGUT}) imply that we can probe the GUT-scale mass
spectrum from the weak-scale parameters ({\it i.e.}, gauge coupling constants
and mass spectrum of the SUSY particles).\footnote{It was claimed in
Ref.~\cite{BH} that the threshold corrections at the GUT-scale is so large that
one cannot predict the SUSY-breaking scale even if measurements on $\alpha_3$
become much more precise. This high sensitivity on GUT-scale mass spectrum
implies that one can probe it through precision measurements on the weak-scale
parameters. Therefore, our result is consistent with their claim.} Especially,
$M_{H_C}$ is determined independently of $M_V$ and $M_{\Sigma}$.
Eq.~(\ref{MGUT}) determines a combination of the vector and adjoint-Higgs
masses $(M_V^2 M_\Sigma)^{1/3}$, and we will call it as ``GUT-scale''
$M_{GUT}=(M_V^2 M_{\Sigma})^{1/3}$, hereafter.\footnote{This ``GUT-scale''
$M_{GUT}$ does not necessarily correspond to the scale where all three gauge
coupling constants meet.}

So far we have assumed a common mass $m_{SUSY}$ for the SUSY particles, but the
mass splitting among the SUSY particles is also important to determine the
GUT-scale mass spectrum. To avoid unnecessary complications, we restrict
ourselves to the minimal supergravity model \cite{nilles}, where the
SUSY-breaking mass parameters at the weak-scale can be determined from a small
number of parameters at the Planck scale, by using the RG equations \cite{DN}.
Therefore, the squark and the slepton masses are given by
\begin{eqnarray}
m_{\tilde{u}}^2&=&m^2 + 6.28M^2 + 0.35 m_Z^2 \cos2\beta_H,\nonumber\\
m_{\tilde{d}}^2&=&m^2 + 6.28M^2 - 0.42 m_Z^2 \cos2\beta_H,\nonumber\\
m_{\tilde{u}^c}^2&=&m^2 + 5.87M^2 + 0.16 m_Z^2 \cos2\beta_H,\nonumber\\
m_{\tilde{d}^c}^2&=&m^2 + 5.82M^2 - 0.08 m_Z^2 \cos2\beta_H,\\
m_{\tilde{e}}^2&=&m^2 + 0.52M^2 - 0.27 m_Z^2 \cos2\beta_H,\nonumber\\
m_{\tilde{\nu}}^2&=&m^2 + 0.52M^2 + 0.50 m_Z^2 \cos2\beta_H,\nonumber\\
m_{\tilde{e}^c}^2&=&m^2 + 0.15M^2 -0.23 m_Z^2 \cos2\beta_H,\nonumber
\label{SCALARMASS}
\end{eqnarray}
in terms of the universal scalar mass $m$ and the gaugino mass $M$ at the
GUT-scale. We have neglected the contributions from the Yukawa couplings to the
renormalization of the particle masses. Also, the gaugino masses at the
weak-scale are given by
\begin{eqnarray}
m_{\tilde{B}} &=& \frac{\alpha_1(m_Z)}{\alpha_5(M_{GUT})}M,\nonumber\\
m_{\tilde{w}} &=& \frac{\alpha_2(m_Z)}{\alpha_5(M_{GUT})}M,\\
m_{\tilde{g}} &=& \frac{\alpha_3(m_Z)}{\alpha_5(M_{GUT})}M,\nonumber
\end{eqnarray}
where $m_{\tilde{B}}$ and $m_{\tilde{g}}$ represent masses of bino and gluino,
respectively.

The effect of the mass splitting can be taken into account by replacing $\ln
(m_{SUSY}/m_Z)$ in Eqs.~(\ref{MHC}), (\ref{MGUT}) as
\begin{eqnarray}
-2 \ln \frac{m_{SUSY}}{m_Z}
	&\longrightarrow& 4 \ln \frac{m_{\tilde{g}}}{m_{\tilde{w}}}
		+\frac{N_g}{5} \ln
		\frac{m_{\tilde{u}^c}^3 m_{\tilde{d}^c}^2 m_{\tilde{e}^c}}
			{m_{\tilde{Q}}^4 m_{\tilde{L}}^2}
%	\nonumber \\
%& &
		-\frac{8}{5} \ln \frac{m_{\tilde{h}}}{m_Z}
		-\frac{2}{5} \ln \frac{m_H}{m_Z}
			\label{SUSY1}
\end{eqnarray}
in Eq.~(\ref{MHC}), and
\begin{equation}
8 \ln \frac{m_{SUSY}}{m_Z}
	\longrightarrow 4 \ln \frac{m_{\tilde{g}}}{m_Z}
		+ 4 \ln \frac{m_{\tilde{w}}}{m_Z}
	+N_g \ln \frac{m_{\tilde{Q}}^2}{m_{\tilde{e}^c} m_{\tilde{u}^c}}
			\label{SUSY2}
\end{equation}
in Eq.~(\ref{MGUT}).  Two doublet Higgs bosons are assumed to have masses at
$m_H$ and $m_Z$, respectively. The symbol $m_{\tilde{h}}$ represents a mass of
doublet Higgsino. We have neglected the mixings among gauginos and doublet
Higgsino.

For the time being we will restrict ourselves to the case where the universal
scalar mass dominates the SUSY breaking ({\it i.e.}, $m\gg M$).  The terms $\ln
(m_{\tilde{u}^c}^3 m_{\tilde{d}^c}^2 m_{\tilde{e}^c}/m_{\tilde{Q}}^4
m_{\tilde{L}}^2)$ in Eq.~(\ref{SUSY1}) and $\ln
( m_{\tilde{Q}}^2/m_{\tilde{e}^c} m_{\tilde{u}^c})$ in Eq.~(\ref{SUSY2}) are
negligibly small. The term $\ln( m_{\tilde{g}}/m_{\tilde{w}})$ stays constant,
since $m_{\tilde{g}}/m_{\tilde{w}} = \alpha_3/\alpha_2 \simeq 3.5$. The
dependence on $m_H$ in Eq.~(\ref{SUSY1}) is weak due to its small coefficient,
and we set $m_H = 1$~\TeV. Therefore, we find that $M_{H_C}$ depends mainly on
the Higgsino mass $m_{\tilde{h}}$, and $M_{GUT}$ on the product of gaugino
masses $m_{\tilde{g}} m_{\tilde{w}}$.  We have also examined the constraint on
$M_{H_C}$ in the no-scale model \cite{noscale} ({\it i.e.}, $m = 0$), and found
that the difference is negligible.

Now we are at the stage to derive the GUT-scale mass spectrum from the above RG
analysis. In our numerical calculation, we use the one-loop RG equations for
the weak- and the GUT-scale thresholds, and the two-loop ones between these two
distant scales. The two-loop RG equations in the minimal SUSY standard model
are \cite{EJ}
\begin{eqnarray}
 \mu \frac{\partial g_i}{\partial \mu} &=&   \frac{1}{16\pi^2} b_i g_i^3
            + \frac{1}{(16\pi^2)^2} \sum_{j=1}^3 b_{ij} g_j^2 g_i^3
\end{eqnarray}
where $i$, $j=1, 2, 3$, and
\begin{eqnarray}
b_i &=&    \left( \begin{array}{c}0 \\  -6\\ -9\end{array}\right)
        +  \left( \begin{array}{c}2 \\   2\\  2\end{array}\right)
           N_g
        +  \left( \begin{array}{c}\frac{3}{10}\\ \frac12\\ 0 \end{array}\right)
           N_{H_f},\label{twoloop}\\
b_{ij}&=&   \left(\begin{array}{ccc}
                 0           &0             &0 \\
                 0           &           -24&0 \\
                 0           &0             &-54
           \end{array}\right)
+          \left(\begin{array}{ccc}
                 \frac{38}{15}&\frac{6}{5}   &\frac{88}{15} \\
                 \frac{2}{5}  &          14  &             8\\
                 \frac{11}{15}&            3 &\frac{68}{3}
           \end{array}\right)N_g  \nonumber\\
&&+        \left(\begin{array}{ccc}
                 \frac{9}{50}&\frac{9}{10}  &0 \\
                 \frac{3}{10}& \frac{7}{2}  &0 \\
                 0           &0             &0
           \end{array}\right)N_{H_f}.
\end{eqnarray}
Here, $N_{H_f}$ is the number of doublet Higgs multiplets ($N_{H_f}=2$).  The
threshold corrections at the two-loop level are expected to be small, since
their mass splittings only within the same order of magnitude do not produce
large logarithms. As the input parameters, we use the $\overline{MS}$ gauge
coupling constants at the $Z$-pole given in Ref.~\cite{LL}, $\alpha = 127.9 \pm
0.2$, $\sin^2 \theta_W = 0.2326 \pm 0.0008$, and $\alpha_3 = 0.118 \pm 0.007$.
However, the use of the simple step-function approximation is only justified in
the $\overline{DR}$-scheme. Since we employ the simple step-function
approximation, we have to convert these coupling constants at the $Z$-pole into
the $\overline{DR}$-scheme by
\begin{equation}
\frac{1}{\alpha_i^{\overline{DR}}}
	=\frac{1}{\alpha_i^{\overline{MS}}}-\frac{C_i}{12\pi} ,
\end{equation}
where $C_1=0$, $C_2=2$, and $C_3=3$ \cite{MSDR}.

Combining all the above discussions, we find that $M_{H_C}$ is constrained to
the range
\begin{equation}
	2.2 \times 10^{13}~\GeV \leq M_{H_C} \leq 2.3 \times 10^{17}~\GeV,
			\label{MHC21}
\end{equation}
and the ``GUT-scale'' is tightly constrained as\footnote{The bound on $M_{GUT}$
quoted in Ref.~\cite{HMY}, $0.90 \times 10^{16}~\GeV \leq M_{GUT} \leq 3.1
\times 10^{16}~\GeV$, includes a minor mistake,
and should be replaced by Eq.~(\ref{MGUT2}).}
\begin{equation}
	0.95 \times 10^{16}~\GeV
	\leq (M_V^2 M_\Sigma)^{1/3} \leq 3.3 \times 10^{16}~\GeV,
			\label{MGUT2}
\end{equation}
for $100~\GeV \leq m_{\tilde{g}} \leq 1~\TeV$. The allowed range of $M_{H_C}$
is much less constrained, because of the small gauge-group representation for
the Higgs multiplets.  The large ambiguity of $M_{H_C}$ comes mainly from the
uncertainty in the strong coupling constant $\alpha_3$, and the prediction on
the nucleon decay will be drastically improved if the uncertainty diminishes.
In Fig.~2 we present the GUT-scale spectrum derived from the present gauge
coupling constants, and also that expected if the error-bar of $\alpha_3$ is
reduced by a factor of 2 with the same central value.  The importance of more
precise measurements on $\alpha_3$ should be clear from the figure.

When one uses these constraints, Eq.~(\ref{MHC21}) and Eq.~(\ref{MGUT2}), one
needs to pay two attentions. First, we have taken only one standard deviation
for gauge coupling constants. If we allow two standard deviations, allowed
region of $M_{H_C}$ spreads to both ends by two orders of magnitude, loosing
any practical limits. On the other hand, the ``GUT-scale'' is still constrained
tightly. Therefore, we restrict our analyses to only one standard deviation as
in Eqs.~(\ref{MHC21}), (\ref{MGUT2}).\footnote{In other words, it is still
possible to raise $M_{H_C}$ up to near the gravitational scale if we allow two
standard deviations. However, we tentatively take this one-sigma bound
seriously to perform further analyses. It should be noted that nucleon decay
can be generated at observable rates for reasonable range of parameters, even
with the color-triplet Higgs of mass at the gravitational scale.} Second, we
have used the RG equations at the two-loop level. One may be concerned for
whether three-loop corrections are important. The difference in $M_{H_C}$
between the one-loop and the two-loop results is a factor of 30, which is a
very small factor compared to the large ratio $M_{GUT}/m_Z$ appearing in the
solutions of the RG equations. Thus we expect that the three-loop corrections
are much less than $O(1)$.

Although we have concentrated to the RG analysis on the gauge coupling
constants to determine the GUT-scale mass spectrum, we will obtain further
constraint on the mass spectrum from the RG analysis on Yukawa coupling
constants. As shown in Eq.~(\ref{MHCmass}), $M_{H_C}$ is given by a Yukawa
coupling constant $\lambda$ between $H_C$, $\overline{H}_C$ and $\Sigma$, which
is not known. On the other hand, the mass $M_V$ is determined by the $SU(5)$
gauge coupling constant $g_5$, whose strength is known by the RG analysis. A
large mass splitting $M_V \ll M_{H_C}$ requires that the $\lambda$ is very
large compared to $g_5$.  Thus the applicability of the perturbation theory
restricts the mass splitting to be not large. The same argument applies to the
mass $M_\Sigma$, which originates in a self-coupling of the adjoint-Higgs as
seen in Eq.~(\ref{24mass}).

A constraint arises by requiring that those Yukawa coupling constants do not
blow-up below the gravitational scale, $M_P/\sqrt{8\pi}=2.4\times 10^{18}$~GeV.
The running of the Yukawa coupling constants in Eq.~(\ref{superpotential}) are
described by the RG equations,
\begin{eqnarray}
\mu \frac{\partial \lambda}{\partial \mu} &=&
	\frac{1}{(4\pi)^2} \left( - \frac{98}{5} g_5^2 + \frac{53}{10}\lambda^2
		+ \frac{21}{40} f^2 + 3 (h^t)^2 \right) \lambda,
%			\nonumber \\
			\\
\mu \frac{\partial f}{\partial \mu} &=&
	\frac{1}{(4\pi)^2} \left( - 30 g_5^2 + \frac{3}{2}\lambda^2
		+ \frac{63}{40} f^2 \right) f,
			\\
\mu \frac{\partial h^t}{\partial \mu} &=&
	\frac{1}{(4\pi)^2} \left( - \frac{96}{5} g_5^2 + \frac{12}{5}\lambda^2
		+ 6 (h^t)^2\right) h^t,
			\\
\mu \frac{\partial g_5}{\partial \mu} &=&
	-\frac{3}{(4\pi)^2} g_5^3,
\end{eqnarray}
where $h^t$ is the Yukawa coupling constant to top quark. The conservative
limit on $\lambda$ can be obtained in the case $f = h^t = 0$. A numerical study
shows that the mass $M_{H_C}$ is limited from above,
\begin{equation}
	M_{H_C} = \frac{\lambda}{\sqrt2g_5} M_V < 2.0 M_V.	\label{MHMV}
\end{equation}
A similar limit on $M_\Sigma$ can be obtained with $\lambda = h^t = 0$,
\begin{equation}
	M_\Sigma = \frac{f}{2\sqrt2g_5} M_V < 1.8 M_V.	\label{MSIGMAMV}
\end{equation}

One may feel uneasy about the assumption that there is no new physics between
the GUT-scale and the gravitational scale. We have examined the above analysis
again requiring that the Yukawa coupling constants do not blow-up below
$10^{17}~\GeV$. This requirement relaxes the constraint Eqs.~(\ref{MHMV}),
(\ref{MSIGMAMV}) at most by a factor of 2.  Therefore, in the following
calculation we use Eqs.~(\ref{MHMV}), (\ref{MSIGMAMV}).

We obtain constraints on $M_V$ and $M_\Sigma$ separately, combining the
discussions above. The limits Eq.~(\ref{MGUT2}) and Eq.~(\ref{MSIGMAMV}) give
\begin{eqnarray}
M_V &>& 0.78 \times 10^{16}~\GeV,	\label{MVlower} \\
M_\Sigma &<& 4.9 \times 10^{16}~\GeV.
\end{eqnarray}
Eq.~(\ref{MVlower}) will be used to put limits on the dimension-six operators
in the following sections.

\section{Present Limits on Dimension-Five Operators}
\setcounter{equation}{0}
In this section we combine the predictions obtained in the previous sections
with the results of the nucleon-decay experiments to see the present status of
the MSGUT.  We find that the present lower bounds on the nucleon partial
lifetimes are still consistent with the SUSY particles below 1~TeV if one
adopts a relatively large value of $M_{H_C}$ ($\ge 2 \times 10^{16} ~\GeV$).
In Table~4, we have listed the experimental lower limits on the partial
lifetimes of nucleon \cite{PDG}.

The most dominant decay mode by dimension-five operators is $n \rightarrow K^0
\bar{\nu}_\mu$, as shown in Table~1. This mode dominates slightly over the
similar decay mode, $p \rightarrow K^+ \bar{\nu}_\mu$, because of the chiral
Lagrangian factor.  These decay modes, however, have an ambiguity in the
parameter $y^{tK}$, which is the relative ratio of the $\tilde{t}$-exchange to
the $\tilde{c}$-exchange contributions. Furthermore, the parameter $y^{tK}$
contains an unknown phase factor $e^{i(\phi_c - \phi_t)}$, which cannot be
determined from any low-energy experiments. In fact, Arnowitt, Chamseddine, and
Nath \cite{AN1} have shown a possible cancellation between the second- and
third-generation contributions. However, if the combination $|1 + y^{tK}|$
decreases, the modes $n,p \rightarrow K \bar{\nu}_\mu$ cease to be dominant.
Then the experimental limit on the other modes $n,p\rightarrow \pi
\bar{\nu}_\mu$ become important. This interchange occurs at
\begin{equation}
		|1+y^{tK}| = 0.40,
\end{equation}
when $|1+y^{t\pi}|=1$. The parameters $y^{tK}$ and $y^{t\pi}$ are correlated as
clear from their definitions (Eqs.~(\ref{ytk}), (\ref{ytpi})). However, we
find it impossible that $|1+y^{tK}|$ and $|1+y^{t\pi}|$ are
canceled out simultaneously, as explained in section~4. We take
$|1+y^{t\pi}|=1$ throughout.

In Fig.~3, we show the lower bound on $M_{H_C}$ derived from the present
nucleon-decay experiments by varying $|1+y^{tK}|$. We choose other parameters
such that the nucleon lifetimes become as long as possible ({\it i.e.},
$m_{\tilde{Q}}=m_{\tilde{L}} = 1~\TeV$, $m_{\tilde{w}} = 45 ~\GeV$, and $\tan
\beta_H=1$). In this figure, the upper horizontal line corresponds to the
maximum value ($M_{H_C} = 2.3\times 10^{17}~\GeV$) in Eq.~(\ref{MHC21}). There
are two curves representing the lower limit on $M_{H_C}$ obtained from the
experimental limits on the nucleon lifetimes. The upper curve corresponds to
the case of the hadron matrix element $\beta = 0.03~\GeV^3$, and the lower
curve to the case of $\beta = 0.003~\GeV^3$. The smaller hadron matrix element
$\beta$ gives weaker constraint as expected. Thus, the conservative lower bound
on $M_{H_C}$ from the nucleon-decay experiments is
\begin{equation}
M_{H_C} \geq 5.3 \times 10^{15} ~\GeV.
\label{MH3EXP}
\end{equation}

We illustrate how the lower bounds on $M_{H_C}$ depend on the SUSY breaking
parameters, the wino mass $m_{\tilde{w}}$ and the squark mass $m_{\tilde{Q}}$
in Fig.~4, assuming $m_{\tilde{L}}\simeq m_{\tilde{Q}}$.\footnote{The situation
does not change even if we allow mass splittings between sleptons and squarks
(say, $m_{\tilde{L}} \ll m_{\tilde{Q}}$), since the denominator in
Eq.~(\ref{squarkheavy}) is dominated by the heavier mass, $m_{\tilde{Q}}$.} In
this figure the dashed line shows the dependence on $m_{\tilde{w}}$ when we
choose the most conservative set of parameters,
$m_{\tilde{Q}}=m_{\tilde{L}}=1~\TeV$, $\tan\beta_H=1$, $|1+y^{tK}|<0.4$, $A_S =
0.67$, and $\beta=0.003~\GeV^3$.  The lower bound on $M_{H_C}$ rises linearly
on $m_{\tilde{w}}$ in the region $m_{\tilde{w}} < 1~\TeV$.  However, the lower
bound decreases as $m_{\tilde{w}}^{-1}$ beyond 1~TeV, and we do not obtain an
upper bound on $m_{\tilde{w}}$ with this conservative choice of parameters.
The dash-dotted line shows the dependence on $m_{\tilde{Q}}$ again for the most
conservative case, $m_{\tilde{w}}=45~\GeV$, $\tan\beta_H=1$, $|1+y^{tK}|<0.4$,
$A_S = 0.67$, and $\beta=0.003~\GeV^3$.  The lower bound on $M_{H_C}$ goes down
as $1/m_{\tilde{Q}}^2$, leading to the lower bound on $m_{\tilde{Q}}$,
\begin{equation}
m_{\tilde{Q}} \ge 150~\GeV.
\end{equation}

We also show the dependence on $\tan\beta_H$ in Fig.~5.  We have fixed
$m_{\tilde{Q}}=m_{\tilde{L}}=1~\TeV$, $m_{\tilde{w}}=45~\GeV$, $\beta=0.003
{}~\GeV^3$, and $|1+y^{tK}|=1.0$ or $|1+y^{tK}|<0.4$.  The lower bound on
$M_{H_C}$ is proportional to
\begin{equation}
 \frac{1}{\sin2\beta_H} =
	\frac12 \left( \frac{1}{\tan\beta_H} + \tan \beta_H \right).
\end{equation}
When $\tan\beta_H \gg 1$, the dependence is almost linear. We find a constraint
on $\tan\beta_H$,
\begin{equation}
\tan\beta_H \le 85,
\end{equation}
which is, however, much weaker than $\tan \beta_H < 40$ obtained from the
requirement that the Yukawa coupling constant $f^b$ for bottom quark
remains in the perturbative regime below the ``GUT-scale''.\footnote{The
$\tan\beta_H$ has to be larger than 0.5, since otherwise the top-quark Yukawa
coupling constant will blow up below the ``GUT-scale'' with $m_t \ge 90~\GeV$.}

Taking $M_{H_C}$ as heavy as possible given in Eq.~(\ref{MHC21}), we obtain
limits on the masses $m_{\tilde{w}}$ and $m_{\tilde{Q}}$. The allowed region is
shown in Fig.~6. Here we have taken $|1+y^{tK}|=1$ and $m_{\tilde{L}} \sim
m_{\tilde{Q}}$. The present experimental limits on the wino and the squark
masses from direct-search experiments at LEP \cite{LEP2} and CDF \cite{CDF} are
shown for comparison.  Since the decay rate behaves like
$(m_{\tilde{w}}/m_{\tilde{Q}}^2)^2$ in the region $m_{\tilde{w}} \ll
m_{\tilde{Q}}$, the lower bound on $m_{\tilde{Q}}$ behaves like
$m_{\tilde{w}}^{1/2}$. In the other extreme, $m_{\tilde{w}} \gg m_{\tilde{Q}}$,
the decay rate goes like $1/m_{\tilde{w}}^2$, and around $m_{\tilde{w}} \sim
10^5$~GeV the constraint on $m_{\tilde{Q}}$ from the nucleon-decay experiments
becomes weaker than that from the CDF experiments.  We see that the ``natural"
mass region $\lsim 1$~TeV for the SUSY particles still survives the
nucleon-decay experiments.

Though the authors of Ref.~\cite{AN2} claimed that the present limits on the
nucleon decay are stringent enough to exclude the SUSY particles lighter than
1~TeV in the absence of a delicate cancellation between matrix elements of the
dimension-five operators ({\it i.e.}, $\beta \simeq 0.003 ~\GeV^3$, $|1+y^{tK}|
\simeq 0.2$), we see now that there is a wide allowed range. This is mainly
because we use the maximum value of $M_{H_C}$ ($2.3\times 10^{17}~\GeV$) given
from the RG analysis while they chose $M_{H_C} \simeq M_{GUT}$ just by hand
({\it i.e.}, $2.0 \times 10^{16} ~\GeV$).

We show similar limits from the $\pi \bar{\nu}_{\mu}$ mode in Fig.~7. As
discussed above, these decay modes become dominant if $|1+y^{tK}| < 0.4$. The
limit is weaker compared to the previous case without the cancellation as shown
in Fig.~6.

We have taken the largest value of $M_{H_C}$ in Eq.~(\ref{MHC21}) in Figs.~6,7.
This value requires $M_V$ larger than $1 \times 10^{17} ~\GeV$ in order to
satisfy the requirement Eq.~(\ref{MHMV}), which leads to $M_{\Sigma}$ smaller
than $3 \times 10^{15} ~\GeV$.  Though this case needs a mass splitting among
the heavy particles, it is still within two orders of magnitude, and it is
completely acceptable phenomenologically.

\setcounter{footnote}{0}
In the minimal supergravity model, the SUSY particle masses are determined
mainly by the universal scalar mass and the gaugino mass. In models where the
universal scalar mass dominates the SUSY breaking parameters, squarks and
sleptons are almost degenerate, that are larger than wino.  Therefore, models
of this type are preferred. In the no-scale model the SUSY breaking is
dominated by the gaugino mass \cite{noscale}, which results in a definite
prediction of the SUSY particle masses ($m_{\tilde{Q}} \sim m_{\tilde{g}}
\simeq 3 m_{\tilde{w}} \sim 3 m_{\tilde{L}}$) \cite{IKYY}.  Since all the mass
parameters in the triangle factor $f$ defined in Eq.~(\ref{fff}) are
proportional to $m_{\tilde{w}}$, $f$ behaves as $m_{\tilde{w}}^{-1}$. This
enables us to derive the lower bound on $m_{\tilde{w}}$ from the nucleon-decay
experiments. We have found that lower limit on $m_{\tilde{w}}$ is 70$~\GeV$
(equivalently, $m_{\tilde{Q}} > 210~\GeV$), in the most conservative case ({\it
i.e.}, $\beta=0.003~\GeV^3$, $M_{H_C}=2.3 \times 10^{17}~\GeV$,
$\tan\beta_H=1$, and $|1+y^{tK}|=0.4$).\footnote{In the absence of the
cancellation ({\it i.e.},\/ $|1+y^{tK}|=1$), the corresponding limit is
$m_{\tilde{w}} > 180~\GeV$.} Thus, we conclude that the no-scale model is still
surviving.\footnote{The wino mass in this region is consistent with the limit
$m_{\tilde{w}} < 300~\GeV$ \cite{IKYY} in the radiative breaking scenario of
the electroweak symmetry in the no-scale model.}

Finally, we briefly discuss the nucleon decay caused by the dimension-six
operators, which come from the $X$ and $Y$ gauge-boson exchanges. We show the
decay rates by the dimension-six operators in Appendix A. We find that
dimension five operators always dominate dimension-six operators, in agreement
with the naive expectations. The ratio ($R_\tau$) of the partial lifetime of
the decay $n
\rightarrow K^0 \bar{\nu}_\mu$ via the dimension-five operators
to that of the decay $p
\rightarrow \pi^0 e^+$ via the dimension-six operators is
\begin{eqnarray}
R_\tau &\equiv&
	\frac{\tau(n \rightarrow K^0 \bar{\nu}_{\mu})}
		{\tau(p \rightarrow \pi^0 e^+)}\label{ratio5to6}\\
	&\leq& 2.8 \times 10^{-4} \times
                \left(\frac{M_{H_C}}{M_V}\right)^4
		\left(\frac{10^{16} ~\GeV}{M_{H_C}}\right)^2, \nonumber
\end{eqnarray}
where all parameters besides $M_{H_C}$ were chosen so that $R_\tau$ becomes as
large as possible.  The ratio of $M_{H_C}$ to $M_V$ is smaller than 2.0 from
Eq.~(\ref{MHMV}), and $R_{\tau}$ is always smaller than $\frac{1}{60}$. The
MSGUT predicts that the dimension-five operators should be observed earlier
than dimension-six operators.

\section{Future Tests on the Minimal SUSY $SU(5)$ GUT}
\setcounter{equation}{0}
Now we examine how stringent constraint on MSGUT we can obtain from the
nucleon-decay experiments in the near future. The superKAMIOKANDE experiment
will push up the lower bound of the nucleon lifetime by a factor of 30.
Meanwhile, the LEP-\uppercase\expandafter{\romannumeral2} experiment will be
able to find wino below $m_Z$. Since the larger $m_{\tilde{w}}$ means faster
nucleon decay as we discussed in previous sections, the combination of these
two experiments can put more stringent limits on the dimension-five operators.

In Figs.~8 and 9, we show the expected limits on the dimension-five operators
if superKAMIOKANDE does not observe the $K\bar{\nu}_\mu$ and $\pi
\bar{\nu}_\mu$ decay modes, and if LEP-\uppercase\expandafter{\romannumeral2}
does not discover wino below $m_Z$. For the $K \bar{\nu}_\mu$ decay modes,
superKAMIOKANDE and LEP-\uppercase\expandafter{\romannumeral2} will be able to
exclude most of the region with the SUSY particles below 1~TeV, even with the
smallest hadron matrix element, $\beta = 0.003~\GeV^3$. In the case where the
$\pi \bar{\nu}_\mu$ mode is dominant, the region below $1 ~\TeV$ is almost
closed leaving a little window. Thus one can see that the
LEP-\uppercase\expandafter{\romannumeral2} and the superKAMIOKANDE experiments
are extremely important for testing the MSGUT.

One may be concerned for a little region below $1 ~\TeV$ which may be left by
these experiments. This window is open because of the large maximum value of
$M_{H_C}$ obtained from the gauge coupling unification. If the error-bar of
$\alpha_3$ is reduced by a factor of 2 with the same central value in the
future, the maximum value of $M_{H_C}$ becomes $6.1 \times 10^{16} ~\GeV$, and
one will be able to close the window completely. In Fig.~10 we demonstrate the
case where the error-bar of $\alpha_3$ is reduced by a factor of 2.  If one
wishes to test the MSGUT completely, it is quite important to reduce the
error-bar of $\alpha_3$.

At the end of this section, we see whether the nucleon decay via the
dimension-six operators can be observed in the near future. From
Eq.~(\ref{MVlower}) and Eq.~(\ref{epi}), the theoretical lower limit of the
partial lifetime of the decay $p \rightarrow \pi^0 e^+$ is obtained as
\begin{equation}
\tau(p \rightarrow \pi^0 e^+) \geq 4.1 \times 10^{33}
	\left( \frac{0.03~\GeV^3}{\alpha} \right)^2  \mbox{years},
\end{equation}
in terms of the hadron matrix element $\alpha$ (=0.003--0.03$~\GeV^3$, see
Appendix A). Since superKAMIOKANDE is expected to reach up to $\tau(p
\rightarrow \pi^0 e^+) \simeq 10^{34}$years \cite{totuka}, there is a
possibility to observe the nucleon decay via the dimension-six operators in the
MSGUT.

\section{Conclusions and Discussions}
\setcounter{equation}{0}
We have analyzed the nucleon decay in the minimal SUSY $SU(5)$ GUT (MSGUT) in
details. First, we have studied the GUT-scale particle spectrum using the RG
analysis, and found a maximum value of $M_{H_C}$ to be $2.3 \times
10^{17}~\GeV$. Then, we have studied the nucleon partial lifetimes with the
largest possible $M_{H_C}$. We have found that the present nucleon-decay
experiments are still consistent with the MSGUT, even without a cancellation
between the matrix elements from exchanges of squarks in different generations.
We have emphasized the important role of precise measurement of the gauge
coupling constants, especially the QCD coupling constant $\alpha_3$ to
determine the mass $M_{H_C}$. We have also stressed that the combined
information of the lower bound on the chargino mass (LEP-II) and on the nucleon
lifetimes (superKAMIOKANDE), will give a stringent constraint on the MSGUT.

It deserves to mention that the nucleon decay prefers supergravity models where
the SUSY breaking parameters are dominated by scalar masses, rather than by
gaugino masses. For example, the no-scale model, where the SUSY breaking
parameters come only from the gaugino masses, is strongly constrained from the
nucleon-decay experiments, though it is still viable.  Increasing the limits on
the gaugino masses will have strong impact on the MSGUT phenomenology.

It is important to see whether predictions on the dimension-five operators
become drastically altered by modifying the MSGUT.  In the MSGUT, there is a
prediction of mass relation, $m_b=m_{\tau}$, $m_s=m_{\mu}$, and $m_d=m_{e}$, at
the GUT-scale. It is known that, though $m_b=m_{\tau}$ is consistent with
observations if top quark is not too heavy \cite{ramond}, $m_s=m_{\mu}$ and
$m_d=m_{e}$ are not. Therefore, the modification of the Yukawa coupling
structure is needed. Georgi and Jarlskog proposed that $m_b=m_{\tau}$,
$3m_s=m_{\mu}$, and $m_d=\frac13 m_{e}$ at the GUT-scale, introducing a
$\mbox{\bf 45}$ dimensional Higgs scalar \cite{GJ}. Also, Kim and \"{O}zer
proposed to make an ``effective'' $\mbox{\bf 45}$ dimensional Higgs scalar by
using higher dimension operators \cite{kim}. These modifications may produce
more uncertainties in the Yukawa couplings of $H_C$ and $\overline{H}_C$ to
matter multiplets.  However, we have checked that these models receive at most
a few times stronger constraint, and hence the main conclusion in the present
analyses does not change qualitatively.

It has been argued that the introduction of a Peccei-Quinn symmetry may be able
to eliminate the dimension-five operators \cite{SYW}.  However, the requirement
of the coupling constant unification does not allow us to introduce new
multiplets of arbitrary masses. We have shown in a separate paper
\cite{PQ} that a suppression of the dimension-five operators
%can be achieved by introducing the Peccei-Quinn symmetry, but that the
%suppression cannot be stronger than in the MSGUT.
by introducing the Peccei-Quinn symmetry cannot be stronger than in the MSGUT.

Another interesting result of our RG analysis on the GUT-scale spectrum is that
one can raise the grand unification scale up to the gravitational scale, by
lowering the $M_{\Sigma}$ down to the intermediate scale ($ 10^{11-12}~\GeV$).
In this case the heavy vector multiplet as well as $H_C$ and $\overline{H}_C$
lie at the gravitational scale, if one allow for two standard deviations in
Eq.~(\ref{MHC21}).  These light $\Sigma_8$ and $\Sigma_3$ require an extremely
small Yukawa coupling $f\Sigma^3$, leading to a very flat potential of
$\Sigma$. The possibility that the three gauge coupling constants meet at the
gravitational scale has come out with a seemingly accidental parameter tuning.
However, this may suggest a completely different underlying physics (non-GUT)
like the superstring theory \cite{GSW}.

\section*{Acknowledgement}
We would like to acknowledge K.~Inoue for useful discussions.
\section*{Note added}
After completing this work, we have received preprints by R.~Arnowitt and
P.~Nath \cite{AN3}. In these works, they also derive a constraint on $M_{H_C}$,
requiring the Yukawa coupling constants to remain in the perturbative regime.
The difference between their constraint $M_{H_C} < 3 M_V$ and our
Eq.~(\ref{MHMV}) is due to different requirements. Namely, they impose that the
Yukawa coupling constant $\lambda$ should not blow up below $2 M_{H_C}$, while
we impose that below the gravitational scale. However, they are not aware of
the possibility to raise $M_V$ by lowering $M_\Sigma$, and hence obtain a
smaller upper bound on $M_{H_C}$. This leads to the opposite conclusion on the
no-scale model, where they claim it to be excluded while we have found it still
viable.

\appendix
\renewcommand{\thesection}{Appendix \Alph{section}}
\renewcommand{\theequation}{\Alph{section}.\arabic{equation}}

\section{Decay Rates via Dimension-six Operators}
\setcounter{equation}{0}
In this appendix we analyze the nucleon-decay rates via dimension-six
operators.  The dimension-six operators are caused by exchanges of $X$ and $Y$
gauge-bosons or color-triplet Higgs scalars.  The color-triplet Higgs scalars
interact with matter only by small Yukawa coupling constants, and also its mass
should be larger than $5 \times 10^{15}~\GeV$ (see Eq.~(\ref{MH3EXP})). Thus
the dimension-six operators induced by the color-triplet Higgs scalar exchanges
are negligible.

The dimension-six operators via the $X$ and $Y$ gauge-boson exchanges are
dominated by generation-diagonal decay modes, and the decay modes containing
strangeness are suppressed by the Cabibbo angle $\sin \theta_C$. We discuss
here only the dominant decay mode $p \rightarrow \pi^0 e^+$. The amplitude by
the $X$ and $Y$ gauge-boson exchanges is
\begin{eqnarray}
{\cal L}_{X,Y} &=& A_R e^{i\phi_u}
	\frac{g_5^2}{M_V^2} \epsilon_{\alpha\beta\gamma}
	\left(  (d_R^{\alpha} u_R^{\beta})(u_L^{\gamma} e_L)
	+ (1+|V_{ud}|^2) (d_L^{\alpha} u_L^{\beta})(u_R^{\gamma} e_R)
                       \right).
\label{dimen6}
\end{eqnarray}
The renormalization factor $A_R$ calculated by the authors in Ref.~\cite{IM} is
$A_R=3.6$. Since the decay rate of this mode caused by the dimension-five
operator is extremely suppressed, the observation of this decay mode would
suggest the presence of the $X$ and $Y$ gauge-bosons exchanges. We calculate
the decay rate using the chiral Lagrangian technique, as
\begin{equation}
  \tau(p \rightarrow \pi^0 e^+)=1.1 \times 10^{36}
   \times \left(\frac{M_V}{10^{16}~\GeV}\right)^4
	\left(\frac{0.003~\GeV^3}{\alpha}\right)^2	\mbox{years}.
			\label{epi}
\end{equation}
Here, $\alpha$ is the hadron matrix element
\begin{equation}
\alpha u_L(\vec{k}) \equiv \epsilon_{\alpha \beta \gamma}
	\langle 0| (d_R^{\alpha} u_R^{\beta}) u_L^{\gamma}
		|p,\vec{k}\rangle,
\label{alpha}
\end{equation}
whose absolute value is the same as $\beta$ ({\it i.e.},\/ $|\alpha| =
|\beta|$) \cite{BEHS}. It is straightforward to compare it with the
nucleon-decay rates via the dimension-five operators. The result is given in
Eq.~(\ref{ratio5to6}).

\section{Renormalization Factors}
\setcounter{equation}{0}
In this appendix, we present several formulae necessary to compute the
renormalization factors which have appeared in the section 3.  There are three
kinds of renormalization effects. First, we have to derive the Yukawa coupling
constants of $H_C$ and $\overline{H}_C$ from the observed quark masses.
Second, the dimension-five operators derived at the GUT-scale receive anomalous
dimensions from the wave-function renormalizations of the external fields.
Third, the four-fermi operators dressed at the SUSY breaking scale should be
renormalized down to 1~GeV. In the text the short-range renormalization factor
between the SUSY breaking and the GUT-scales is denoted as $A_S$, and the
long-range renormalization factor between 1~GeV and the SUSY breaking scale as
$A_L$. Though $A_S$ and $A_L$ are estimated by the authors in Ref.~\cite{ENR},
we need minor corrections to their calculation of $A_S$. We demonstrate the
derivation of $A_S$, and also comment on $A_L$.

Since the Yukawa coupling constants in Eq.~(\ref{dimen5}) are those given at
the GUT-scale, we have to calculate the values of the Yukawa coupling constants
by solving the RG equations from the SUSY breaking scale up to the GUT-scale.
This was done in the Ref.~\cite{ENR}. Since the Yukawa couplings $H_f$ and
$\overline{H}_f$ are $F$-terms, what we have to compute is only the
wave-function renormalizations of each chiral superfields thanks to the
non-renormalization theorem of the $F$-terms \cite{fterm}.  The authors of
Ref.~\cite{inoue} have given the following general formula for the running of
arbitrary $F$-terms. Any $F$-terms can be re-written in the following form,
\begin{equation}
	W = \frac{1}{3!} h^{klm} \phi_k \phi_l \phi_m,
\end{equation}
where $\phi$'s are chiral superfields, and the coupling constant $h^{klm}$ are
supposed to be completely symmetric under the interchange of the indices
$k,\,l,\,m$. If the fields $\phi_k$'s belong to the representation $R^k$ of
certain gauge group $i$, the coupling constants $h^{klm}$ follow the RG
equation
\begin{equation}
	\frac{{\rm d} h^{klm}}{{\rm d} \ln \mu}
		= \frac{1}{(4\pi)^2} \left[
			\theta^k_{k'} h^{k' lm} + \theta^l_{l'} h^{kl' m}
			+ \theta^m_{m'} h^{klm'} \right],
\end{equation}
where the constants $\theta$'s are given by
\begin{equation}
	\theta^k_{k'} = -2 C_2^i (R^k) g_i^2 \delta^k_{k'}
		+ \frac{1}{2!} h^{kpq} h^\ast_{k'pq}.
\end{equation}
The gauge coupling constant is denoted by $g_i$, and the second Casimir of the
representation $R^k$ by $C_2^i (R^k)$. If the coupling constant $h^{klm}$ can
be neglected in the expression of $\theta$, the RG equation can be easily
integrated to the form
\begin{equation}
	h^{klm} (\mu) = h^{klm} (\mu_0)
                \prod_{r=k,l,m}
		\prod_{i=1,2,3} \left(
			\frac{\alpha_i (\mu_0)}{\alpha_i (\mu)}
			\right)^{C_2^i(R^r)/b_i}.
\end{equation}
Here, the coefficient of the $\beta$-function $b_i$ is defined as
\begin{equation}
	\frac{{\rm d}\alpha_i^{-1}}{{\rm d} \ln \mu}
		= -\frac{b_i}{2\pi}.
\end{equation}
The explicit forms of $b_i$'s in the minimal SUSY standard model are given in
Eq.~(\ref{twoloop}).  Thus, in our case, the Yukawa coupling constants of the
lower generations will be renormalized from the SUSY breaking to the GUT-scales
by
\begin{eqnarray}
	h^{u} (M_{GUT}) &=& h^u (m_Z)
			\left( \frac{\alpha_1 (m_Z)}{\alpha_{5}(M_{GUT})}
			\right)^{13/198}
			\left( \frac{\alpha_2 (m_Z)}{\alpha_{5}(M_{GUT})}
			\right)^{3/2}
			\left( \frac{\alpha_3 (m_Z)}{\alpha_{5}(M_{GUT})}
			\right)^{-8/9}, \nonumber \\
	f^{d} (M_{GUT}) &=& f^d (m_Z)
			\left( \frac{\alpha_1 (m_Z)}{\alpha_{5}(M_{GUT})}
			\right)^{7/198}
			\left( \frac{\alpha_2 (m_Z)}{\alpha_{5}(M_{GUT})}
			\right)^{3/2}
			\left( \frac{\alpha_3 (m_Z)}{\alpha_{5}(M_{GUT})}
			\right)^{-8/9}, \\
	f^{e} (M_{GUT}) &=& f^e (m_Z)
			\left( \frac{\alpha_1 (m_Z)}{\alpha_{5}(M_{GUT})}
			\right)^{3/22}
			\left( \frac{\alpha_2 (m_Z)}{\alpha_{5}(M_{GUT})}
			\right)^{3/2}. \nonumber
\end{eqnarray}
We have assumed the SUSY breaking scale to be $m_Z$. Here, $h^u,\,f^d,\,f^e$
are the Yukawa coupling constants of $H_f$ and $\overline{H}_f$ to up-,
down-type quark and charged-lepton multiplets. Note that the expressions differ
from those given in Ref.~\cite{ENR}.\footnote{The authors in Ref.~\cite{ENR}
seem to have considered the renormalization of the mass operators rather than
the Yukawa coupling constants. Since what determines the coefficient of the
dimension-five operators is the Yukawa coupling constants themselves rather
than the mass parameters induced by the $SU(2) \times U(1)$ breaking, it is
clear that one should consider the running of the Yukawa coupling constants up
to the GUT-scale.} The most important difference from the previous analysis is
that the formula in Ref.~\cite{ENR} does not have the factors of $\alpha_2$,
and hence they overestimated the Yukawa coupling constants at the GUT-scale.

Another ingredient which has not been included in Ref.~\cite{ENR} is the large
top-quark Yukawa coupling constant. Its effect appears in two points. First,
the large Yukawa coupling constant contributes to the wave-function
renormalization of the Higgs multiplet. If top quark becomes heavy, its
large Yukawa coupling constant cannot be neglected in the RG equation of the
Yukawa coupling constants of first and second generations. Its effect is to
enhance the Yukawa coupling constants at the GUT-scale , and hence enhancing
the dimension-five operators. Second, the top-quark Yukawa coupling constant
itself become quite large at the GUT-scale, due to the wave-function
renormalization of the top-quark multiplet. This further enhances the
dimension-five operators for the third-generation contribution. Since the RG
equation cannot be solved analytically if one includes the top-quark Yukawa
coupling constant, we can only present the numerical results.

{}From the Yukawa coupling constants at the GUT-scale,
we obtain the coefficient
of the dimension-five operators in Eq.~(\ref{dimen5}). The dimension-five
operators receive the anomalous dimensions, which we have to estimate to know
their coefficients at the SUSY breaking scale. Here again, since the
dimension-five operators are $F$-terms, all the renormalization effects come
from the wave-function renormalizations of the external lines. We obtain the
same result as that given in Ref.~\cite{ENR}. The $QQQL$ operators including
only first- and second-generation fields in Eq.~(\ref{dimen5}) receive an
enhancement factor
\begin{equation}
	\left( \frac{\alpha_1 (m_Z)}{\alpha_{5}(M_{GUT})} \right)^{-1/33}
	\left( \frac{\alpha_2 (m_Z)}{\alpha_{5}(M_{GUT})} \right)^{-3}
	\left( \frac{\alpha_3 (m_Z)}{\alpha_{5}(M_{GUT})} \right)^{4/3}.
\end{equation}
The enhancement factor of the operators including the third-generation fields
is a little reduced by the effect of the top-quark Yukawa coupling.

We  show the numerical values of the
coefficient $A_S(i,j,k)$,  the short-range
renormalization effect between the GUT- and the SUSY breaking scales.
First, if
the contribution of the top-quark Yukawa coupling constant is neglected,
this value becomes
\begin{eqnarray}
A_S
&=&
\left( \frac{\alpha_1(m_Z)}{\alpha_5(M_{GUT})} \right)^{{7}/{99}}
\left( \frac{\alpha_3(m_Z)}{\alpha_{5}(M_{GUT})} \right)^{-4/9}
        \nonumber \\
&=&0.59.
\end{eqnarray}
Thus our short range renormalization factor $A_S$ is smaller than that in
Ref.~\cite{ENR} by $\frac23$ (their value is 0.91).  Next, we show the
numerical values of the coefficient $A_S$ in Fig.~11 for varying
$m_t/\sqrt2\sin\beta_H$. Since $A_S$ depend on the top-quark Yukawa coupling
constant rather than $m_t$, only a combination $m_t/\sqrt2\sin\beta_H$ is
relevant.  The solid line represents $A_S$ for the dimension-five operators
only with first- and second-generation fields, and the dash-dotted line for
the operator $(Q_t Q_t) (Q_c L_\mu)$. The lower horizontal line is $A_S$ with
the top-quark contribution neglected.  One can see that the top-quark Yukawa
coupling enhances the dimension-five operators.

The factor $A_L$ in Eq.~(\ref{dim6}) is the long-range renormalization factor
due to the QCD interaction between the SUSY scale and 1~GeV scale, and contain
the renormalization of Yukawa coupling constants and the anomalous dimension of
four fermi-operators. It is given in Ref.~\cite{ENR},
\begin{eqnarray}
A_L&=&\left(\frac{\alpha_3(1~\GeV)}{\alpha_3(m_c)}\right)^{-2/3}
    \left(\frac{\alpha_3(m_c)}{\alpha_3(m_b)}\right)^{-{18}/{25}}
\left(\frac{\alpha_3(m_b)}{\alpha_3(m_Z)}\right)^{-{18}/{23}}\nonumber\\
&=&0.22
\end{eqnarray}
Combining all the renormalization effects, the four-fermi operators can be
written down as in Eq.~(\ref{dim6}).

\section{Chiral Lagrangian Technique}
\setcounter{equation}{0}
In this appendix we present a chiral Lagrangian technique for translating
operators at the quark level to those at the hadron level. This technique has
been developed in the Refs.~\cite{CWH,cd}

The baryon number violating four-fermi operators derived from the
dimension-five operators are
\begin{eqnarray}
{O}(duu\nu_i) &=& \epsilon_{\alpha\beta\gamma}
                    (d_L^{\alpha} u_L^{\beta})(d_L^{\gamma} \nu_{iL}),
                     \nonumber\\
{O}(dude_i) &=& \epsilon_{\alpha\beta\gamma}
                    (d_L^{\alpha} u_L^{\beta})(u_L^{\gamma} e_{iL}),
                     \nonumber\\
{O}(sud\nu_i) &=& \epsilon_{\alpha\beta\gamma}
                    (s_L^{\alpha} u_L^{\beta})(d_L^{\gamma} \nu_{iL}),
                     \\
{O}(dus\nu_i) &=& \epsilon_{\alpha\beta\gamma}
                    (d_L^{\alpha} u_L^{\beta})(s_L^{\gamma} \nu_{iL}),
                     \nonumber\\
{O}(suue_i) &=& \epsilon_{\alpha\beta\gamma}
                    (s_L^{\alpha} u_L^{\beta})(u_L^{\gamma} e_{iL}),
                     \nonumber
\end{eqnarray}
where quark and lepton fields are in mass eigenstates, and $i$ denotes the
generation indices.  There are also baryon number violating four-fermi
operators derived from the dimension-six operators, and we will concentrate
only on the operators relevant for the decay mode $p \rightarrow \pi^0 e^+$,
\begin{eqnarray}
\tilde{O}^{(1)} &=& \epsilon_{\alpha\beta\gamma}
                        (d_R^{\alpha} u_R^{\beta})(u_L^{\gamma} e_{L}),
                        \nonumber\\
\tilde{O}^{(2)} &=& \epsilon_{\alpha\beta\gamma}
                        (d_L^{\alpha} u_L^{\beta})(u_R^{\gamma} e_{R}).
\end{eqnarray}
The effective Lagrangian ${\cal{L}}^q$ at the quark level for each decay modes
can be written as
\begin{eqnarray}
{\cal L}^q(n,p \rightarrow \pi (\eta) \bar{\nu}_i) &=& C(duu\nu_i) O(duu\nu_i),
               \nonumber\\
{\cal L}^q(n,p \rightarrow \pi (\eta)  e_i^+) &=& C(dude_i) O(dude_i),
               \nonumber\\
{\cal L}^q(n,p \rightarrow K \bar{\nu}_i) &=& C(sud\nu_i) O(sud\nu_i)
                                          +  C(dus\nu_i) O(dus\nu_i),
               \nonumber\\
{\cal L}^q(n,p \rightarrow K e_i^+) &=& C(suue_i) O(suue_i),
               \nonumber\\
{\cal L}^q(n,p \rightarrow \pi e_i^+) &=& \tilde{C}^{(1)} \tilde{O}^{(1)}
                                    + \tilde{C}^{(2)} \tilde{O}^{(2)}.
\end{eqnarray}
The coefficients $C$'s are derived from Eq.~(\ref{dim6}), and $\tilde{C}$'s
from Eq.~(\ref{dimen6}),
\begin{eqnarray}
C(duu\nu_i) &=& \frac{4\alpha_2^2}{M_{H_C}}
	\frac{\overline{m}_c \overline{m}_{d_i} e^{i\phi_c}
	V_{ud_i}^\ast V_{cd}^2 A_L A_S(c,u,d_i)}{m_W^2 \sin 2\beta_H}
         (1+y^{t\pi})(f(u,\,d) + f(u,\,e)), \nonumber\\
C(dude_i) &=& \frac{2\alpha_2^2}{M_{H_C}}
	\frac{\overline{m}_u \overline{m}_{d_i} e^{i\phi_u}
	V_{cd_i}^\ast V_{cd} A_L A_S(u,c,d_i)}{m_W^2 \sin 2\beta_H}
         (f(u,\,d) + f(d,\,\nu)), \nonumber\\
C(sud\nu_i) &=& C(dus\nu_i) \nonumber\\
            &=& \frac{2\alpha_2^2}{M_{H_C}}
	\frac{\overline{m}_c \overline{m}_{d_i} e^{i\phi_c}
	V_{ud_i}^\ast V_{cd} V_{cs} A_L A_S(c,u,d_i)}{m_W^2 \sin 2\beta_H}
         (1+y^{tK})(f(u,\,d) + f(u,\,e)), \nonumber\\
C(suue_i) &=& \frac{2\alpha_2^2}{M_{H_C}}
	\frac{\overline{m}_u \overline{m}_{d_i} e^{i\phi_u}
	V_{cd_i}^\ast V_{cs} A_L A_S(u,c,d_i)}{m_W^2 \sin 2\beta_H}
         (f(u,\,d) + f(d,\,\nu)), \nonumber\\
\tilde{C}^{(1)} &=& e^{i\phi_u} \frac{g_5^2 A_R}{ M_V^2}, \nonumber\\
\tilde{C}^{(2)} &=& e^{i\phi_u} \frac{g_5^2 A_R}{ M_V^2}(1+|V_{ud}|^2).
\label{ccc}
\end{eqnarray}
We need to translate these effective Lagrangians at the quark level ${\cal
L}^q$ to the operators written in terms of the baryon and meson fields ${\cal
L}^h$ .

Let us review the chiral Lagrangian for baryons and mesons.  The
Nambu-Goldstone bosons $\Phi$ associated with the spontaneous breaking of
chiral $SU(3)_L \times SU(3)_R$ symmetry can be written by
\begin{eqnarray}
U &=& \exp\left(\frac{2i\Phi}{f_{\pi}}\right)
\end{eqnarray}
where $f_{\pi}$ is the pion decay constant, and
\begin{equation}
\Phi=\left(
   \begin{array}{ccc}
    \sqrt{\frac12} \pi^0 + \sqrt{\frac16} \eta & \pi^+ & K^+\\
    \pi^- &  -\sqrt{\frac12} \pi^0 + \sqrt{\frac16} \eta & K^0 \\
    K^-  &  \bar{K}^0 & -\sqrt{\frac23} \eta
   \end{array}
  \right).
\end{equation}
Similarly, the baryon fields can be written in the matrix form,
\begin{equation}
B=\left(
   \begin{array}{ccc}
    \sqrt{\frac12} \Sigma^0 + \sqrt{\frac16} \Lambda & \Sigma^+ & p^+\\
    \Sigma^-  &  -\sqrt{\frac12} \Sigma^0 + \sqrt{\frac16} \Lambda & n^0 \\
    \Xi^-  &  \Xi^0 & -\sqrt{\frac23} \Lambda
   \end{array}
  \right).
\end{equation}
Now the most general $SU(3)_L\times SU(3)_R$ invariant Lagrangian for strong
interactions of mesons and baryons is
\begin{eqnarray}
{\cal L}_0 &=& \frac18 f_{\pi}^2 {\rm Tr} (\partial U)(\partial U^{\dagger})
               + {\rm Tr} \bar{B}(i\!\not\!\partial - M_{inv})B \nonumber\\
          && + \frac12 i {\rm Tr} \bar{B} \gamma^{\mu}
              \left[ \zeta (\partial_{\mu} \zeta^{\dagger})
		+\zeta^{\dagger}(\partial_{\mu} \zeta)\right]B\nonumber\\
          && +\frac12 i {\rm Tr} \bar{B} \gamma^{\mu}B
              \left[ (\partial_{\mu} \zeta)\zeta^{\dagger}
		+ (\partial_{\mu} \zeta^{\dagger})\zeta \right]\nonumber\\
          && -\frac12 i (D-F) {\rm Tr}\bar{B} \gamma^{\mu} \gamma_5 B
              \left[(\partial_{\mu} \zeta)\zeta^{\dagger}
		- (\partial_{\mu} \zeta^{\dagger})\zeta \right]\nonumber\\
          && +\frac12 i (D+F) {\rm Tr}\bar{B} \gamma^{\mu} \gamma_5
              \left[\zeta(\partial_{\mu} \zeta^{\dagger})
		- \zeta^{\dagger}(\partial_{\mu} \zeta) \right] B,
\end{eqnarray}
where
\begin{equation}
\zeta = \exp\left[ \frac{iM}{f_{\pi}}\right].
\end{equation}
Since the (current) quark masses that break the chiral symmetry are small for
up, down, and strange quarks, we can use ${\cal L}_0$ to estimate the lifetimes
of nucleon.

Now we translate the effective Lagrangians containing quark fields ${\cal L}^q$
to the ones containing baryons and mesons ${\cal L}^h$, by comparing the
transformation properties under the $SU(3)_L \times SU(3)_R$ symmetry. The
transformation properties of the baryon number violating operators given above
are,
\begin{eqnarray}
O(duu\nu_i), O(dude_i) &{\rm as}& ({\bf 8,1}),\nonumber\\
O(sud\nu_i), O(suue_i) &{\rm as}& ({\bf 8,1}),\nonumber\\
O(sud\nu_i)                &{\rm as}& ({\bf 8,1}),\nonumber\\
\tilde{O}^{(1)}              &{\rm as}& ({\bf 3,{3}^*}),\nonumber\\
\tilde{O}^{(2)}              &{\rm as}& ({\bf {3}^*,3}).
\end{eqnarray}
Thus, the four-fermi operators translating as ({\bf 8,1}) can be expressed in
terms of the baryon and meson fields with a dimensionful constant $\beta$,
\begin{eqnarray}
{\cal L}^h(n,p \rightarrow \pi (\eta) \bar{\nu}_i)
&=& \beta C(duu\nu_i) \nu_{dL}
               {\rm Tr}[ P_1 \zeta B_L \zeta^{\dagger}] + h.c.,
               \nonumber\\
{\cal L}^h(n,p \rightarrow \pi (\eta)  e_i^+) &=& \beta C(dude_i) e_{dL}
               {\rm Tr}[ P_2 \zeta B_L \zeta^{\dagger}] + h.c.,
               \nonumber\\
{\cal L}^h(n,p \rightarrow K \bar{\nu}_i) &=& \beta C(sud\nu_i) \nu_{dL}
               {\rm Tr}[ P_3 \zeta B_L \zeta^{\dagger}]
                                          + \beta  C(dus\nu_i) \nu_{dL}
               {\rm Tr}[ P_4 \zeta B_L \zeta^{\dagger}] + h.c.,
               \nonumber\\
{\cal L}^h(n,p \rightarrow K e_i^+) &=& \beta C(suue_i) e_{dL}
               {\rm Tr}[ P_5 \zeta B_L \zeta^{\dagger}] + h.c,
\end{eqnarray}
and the ones translating as (${\bf 3,3^*}$) or (${\bf 3^*,3}$) with a
dimensionful constant $\alpha$,
\begin{eqnarray}
{\cal L}^h(n,p \rightarrow \pi e^+) &=& \alpha \tilde{C}^{(1)} e_{dL}
               {\rm Tr}[ P_6 \zeta B_L \zeta]
                                    + \alpha \tilde{C}^{(2)} e_{dR}
               {\rm Tr}[ P_7 \zeta^{\dagger} B_R \zeta^{\dagger}] + h.c,
\end{eqnarray}
with the coefficients $C$'s and $\tilde{C}$'s defined in Eq.~(\ref{ccc}).  In
the above formulae, $P_i$ are projection operators,
\begin{eqnarray}
P_1 = \left(\begin{array}{ccc}
         0&0&0\\0&0&0\\0&1&0
        \end{array} \right),&&
P_2 = \left(\begin{array}{ccc}
         0&0&0\\0&0&0\\1&0&0
        \end{array} \right),\nonumber\\
P_3 = \left(\begin{array}{ccc}
         0&0&0\\0&-1&0\\0&0&0
        \end{array} \right),&&
P_4 = \left(\begin{array}{ccc}
         0&0&0\\0&0&0\\0&0&1
         \end{array} \right),\nonumber\\
P_5 = \left(\begin{array}{ccc}
         0&0&0\\-1&0&0\\0&0&0
         \end{array} \right),&&
P_6 = \left(\begin{array}{ccc}
          0&0&0\\0&0&0\\1&0&0
         \end{array} \right),\nonumber\\
P_7 &=& \left(\begin{array}{ccc}
          0&0&0\\0&0&0\\1&0&0
          \end{array} \right).
\end{eqnarray}
The constants $\alpha$ and $\beta$ are the same as those defined in
Eqs.~(\ref{alpha}), (\ref{beta}) \cite{CWH}.
When one estimates the decay rates of the nucleons, one needs to include the
virtual baryon exchanges.  For example, to estimate the lifetimes of
decay modes $n, p \rightarrow K \bar{\nu}$, one should add the
contributions from diagrams with virtual $\Sigma$ and $\Lambda$ exchanges. We
show the chiral Lagrangian factors of each decay modes in Table~5.
We took $m_B
\equiv m_{\Sigma}=m_{\Lambda}=1150~\MeV$, $D=0.81$, $F=0.44$ in Tables~1--3.

Here we have some comments on the chiral Lagrangian factors in the  decay rates
of the nucleons.  First, the ratio of $K \bar{\nu}$ decay rates of
neutron and proton
is
\begin{eqnarray}
\frac{\Gamma(n \rightarrow K^0 \bar{\nu})}{\Gamma(p \rightarrow K^+ \bar{\nu})}
&=&\frac{\left| 2+\frac{2m_n}{m_B}F \right|^2}
        {\left| 1+\frac{m_p}{m_B} (D+F)\right|^2}\nonumber\\
&=&1.8.
\end{eqnarray}
This shows that the decay rate of
the neutron is larger than that of the proton.
The situation is different in $\pi
\bar{\nu}$ mode. The decay rate of the mode
$p \rightarrow \pi^+ \bar{\nu}$ is two times larger
than that of $n \rightarrow \pi^0 \bar{\nu}$.
However, we use $n \rightarrow \pi^0 \bar{\nu}$
in section 5 because the
experimental lower bound on $\tau(n \rightarrow \pi^0 \bar{\nu})$
is longer than that on $\tau(p \rightarrow \pi^+ \bar{\nu})$.

Second, the decay rates into $\eta$ of dimension-five operators are as large
as those into $\pi^0$.  For example, the ratio in $\bar{\nu}$ decay mode is
\begin{eqnarray}
\frac{\Gamma(n \rightarrow \eta \bar{\nu})}
	{\Gamma(n \rightarrow \pi^0 \bar{\nu})}
&=&\frac{(m_n^2-m_{\eta}^2)^2}{m_n^4}
\frac{3\left| 1-\frac13(D-3F) \right|^2}
        {\left| 1+D+F\right|^2}\nonumber\\
&=&0.35.
\end{eqnarray}
Recall that  the decay rates into $\eta$
from dimension-six operators is negligible \cite{CWH}.
Thus this mode may be interesting.

The results in this appendix depend on the parameters $\alpha$ and $\beta$.
However, they
are sensitive to hadron dynamics, and they differ for each
hadron models. Even among the lattice calculations,
the results vary from $0.03~\GeV^3$ \cite{hara} to
$0.0056~\GeV^3$ \cite{GKSM}.

\newpage

\section*{\large\bf Table 1}
The prediction of the nucleon partial lifetimes for the
dominant decay modes, arising from the dimension-five operators $(Q_c Q_c) (Q_u
L_\mu)$ and $(Q_t Q_t) (Q_u L_\mu)$. This class of decay modes depends on the
parameter $y^{tK}$. The mass degeneracy $m_{\tilde{c}} = m_{\tilde{u}}$ and
$m_{\tilde{\mu}} = m_{\tilde{e}}$ is assumed. The function $f$ is defined in
Eq.~(\ref{fff}) and depends on the SUSY particle masses. See the text for other
variables.
\small
\begin{displaymath}
\begin{array}{ccc}
	\tau(p \rightarrow K^+ \bar{\nu}_\mu) & = & 6.9 \times 10^{31} \\
	\tau(p \rightarrow K^+ \bar{\nu}_e)  & = & 1.4 \times 10^{33} \\
	\tau(n \rightarrow K^0 \bar{\nu}_\mu) & = & 3.9 \times 10^{31} \\
	\tau(n \rightarrow K^0 \bar{\nu}_e)  & = & 7.7 \times 10^{32}
 \end{array}
	\times
	\left| 	\frac{\mbox{0.003~GeV}^3}{\beta}
		\frac{0.67}{A_S} \frac{\sin 2\beta_H}{1+y^{tK}}
		\frac{M_{H_C}}{10^{17}\mbox{~GeV}}
		\frac{\mbox{TeV}^{-1}}{f(u,\,d) + f(u,\,e)} \right|^2
	\mbox{yrs}
\end{displaymath}
\normalsize

\section*{\large\bf Table 2}
The prediction of the nucleon partial lifetimes for the next-leading
decay modes, arising from the dimension-five operators $(Q_c Q_c) (Q_u
L_\mu)$ and $(Q_t Q_t) (Q_u L_\mu)$. This class of modes depends on the
parameter $y^{t\pi}$. The mass degeneracy $m_{\tilde{c}} = m_{\tilde{u}}$ and
$m_{\tilde{\mu}} = m_{\tilde{e}}$ is assumed. The function $f$ is defined in
Eq.~(\ref{fff}) and depends on the SUSY particle masses. See the text for other
variables.
\small
\begin{displaymath}
\begin{array}{ccc}
	\tau(p \rightarrow \pi^+ \bar{\nu}_\mu) & = & 1.4 \times 10^{32} \\
	\tau(p \rightarrow \pi^+ \bar{\nu}_e)  & = & 2.9 \times 10^{33} \\
	\tau(n \rightarrow \pi^0 \bar{\nu}_\mu) & = & 2.9 \times 10^{32} \\
	\tau(n \rightarrow \pi^0 \bar{\nu}_e)  & = & 5.7 \times 10^{33} \\
	\tau(n \rightarrow \eta^0 \bar{\nu}_\mu) & = & 8.2 \times 10^{32} \\
	\tau(n \rightarrow \eta^0 \bar{\nu}_e)  & = & 1.6 \times 10^{34}
	 \end{array}
\times
	\left| 	\frac{\mbox{0.003~GeV}^3}{\beta}
		\frac{0.67}{A_S} \frac{\sin 2\beta_H}{1+y^{t\pi}}
		\frac{M_{H_C}}{10^{17}\mbox{~GeV}}
		\frac{\mbox{TeV}^{-1}}{f(u,\,d) + f(u,\,e)} \right|^2
	\mbox{yrs}
\end{displaymath}
\normalsize

\newpage
\section*{\large\bf Table 3}
The prediction of the nucleon partial lifetimes for the decay modes which
depend neither on the parameter $y^{tK}$ nor $y^{t\pi}$. The relevant
dimension-five operator is $(Q_u Q_u) (Q_c L_\mu)$ alone.  The mass degeneracy
$m_{\tilde{c}} = m_{\tilde{u}}$ and $m_{\tilde{\nu}_{\mu}} =
m_{\tilde{\nu}_{e}}$ is assumed. The function $f$ is defined in Eq.~(\ref{fff})
and depends on the SUSY particle masses. See the text for other variables.
\small
\begin{displaymath}
\begin{array}{ccc}
	\tau(p \rightarrow K^0 \mu^+) & = & 1.0 \times 10^{35} \\
	\tau(p \rightarrow \pi^0 \mu^+)  & = & 2.0 \times 10^{35} \\
	\tau(p \rightarrow \eta^0 \mu^+) & = & 5.7 \times 10^{35} \\
	\tau(n \rightarrow \pi^- \mu^+)  & = & 9.9 \times 10^{34}
	 \end{array}
\times	\left| 	\frac{\mbox{0.003~GeV}^3}{\beta}
		\frac{0.67}{A_S} \sin 2\beta_H
		\frac{M_{H_C}}{10^{17}\mbox{~GeV}}
		\frac{\mbox{TeV}^{-1}}{f(u,\,d) + f(d,\,\nu)} \right|^2
	\mbox{yrs}
\end{displaymath}
\normalsize

\section*{\large\bf Table 4}
 The experimental lower bounds on the nucleon partial
lifetimes at the 90\% C.L. \cite{PDG}.
\small
\begin{displaymath}
\begin{array}{ccc}
	\tau(p \rightarrow K^+ \bar{\nu}) &>& 1.0 \times 10^{32} \mbox{yrs}\\
	\tau(n \rightarrow K^0 \bar{\nu}) &>& 8.6 \times 10^{31} \mbox{yrs}\\
	\tau(p \rightarrow \pi^+ \bar{\nu}) &>& 2.5 \times 10^{31} \mbox{yrs}\\
	\tau(n \rightarrow \pi^0 \bar{\nu}) &>& 1.0 \times 10^{32} \mbox{yrs}\\
	\tau(n \rightarrow \eta \bar{\nu}) &>& 5.4 \times 10^{31} \mbox{yrs}\\
	\tau(p \rightarrow K^0 \mu^+) &>& 1.2 \times 10^{32} \mbox{yrs}\\
	\tau(p \rightarrow \pi^0 \mu^+) &>& 2.7 \times 10^{32} \mbox{yrs}\\
	\tau(p \rightarrow \eta \mu^+) &>& 6.9 \times 10^{31} \mbox{yrs}\\
	\tau(n \rightarrow \pi^- \mu^+) &>& 1.0 \times 10^{32} \mbox{yrs}\\
	\tau(p \rightarrow \pi^0 e^+) &>& 5.5 \times 10^{32} \mbox{yrs}
 \end{array}
\end{displaymath}
\normalsize

\newpage
\section*{\large\bf Table 5}
Chiral Lagrangian factors in the nucleon-decay matrix elements.
For notations, see the text.
\small
\begin{eqnarray*}
\Gamma(p \rightarrow K^+\bar{\nu}_i)
            &=& \frac{(m_p^2-m_K^2)^2}{32\pi m_p^3 f_{\pi}^2}
                \left|C(sud\nu_i) \frac{2m_p}{3m_B}D +
                      C(dus\nu_i) \left[1+\frac{m_p}{3m_B}(D+3F)\right]
		\right|^2\\
\Gamma(n \rightarrow K^0\bar{\nu}_i)
            &=& \frac{(m_p^2-m_K^2)^2}{32\pi m_p^3 f_{\pi}^2}
                \left|C(sud\nu_i) \left[1-\frac{m_n}{3m_B}(D-3F)\right] +
                      C(dus\nu_i) \left[1+\frac{m_p}{3m_B}(D+3F)\right]
		\right|^2\\
\Gamma(p \rightarrow \pi^+\bar{\nu}_i)
            &=& \frac{m_p}{32\pi f_{\pi}^2}
                \left|C(duu\nu_i) \left[ 1+D+F \right] \right|^2\\
\Gamma(n \rightarrow \pi^0\bar{\nu}_i)
            &=& \frac{m_n}{64\pi f_{\pi}^2}
                \left|C(duu\nu_i) \left[ 1+D+F \right] \right|^2\\
\Gamma(n \rightarrow \eta\bar{\nu}_i)
            &=& \frac{(m_n^2-m_\eta^2)^2}{64\pi m_n^3 f_{\pi}^2}
               3 \left|C(duu\nu_i) \left[ 1-\frac13 (D-3F) \right] \right|^2\\
\Gamma(p \rightarrow K^0 e_i^+)
            &=& \frac{(m_p^2-m_K^2)^2}{32\pi m_p^3 f_{\pi}^2}
                \left|C(suue_i) \left[1-\frac{m_p}{m_B}(D-F)\right] \right|^2\\
\Gamma(p \rightarrow \pi^0 e_i^+)
            &=& \frac{m_p}{64\pi f_{\pi}^2}
                \left|C(dude_i) \left[ 1+D+F \right] \right|^2\\
\Gamma(p \rightarrow \eta e_i^+)
            &=& \frac{(m_p^2-m_\eta^2)^2}{64\pi m_p^3 f_{\pi}^2}
               3 \left|C(dude_i) \left[ 1-\frac13 (D-3F) \right] \right|^2\\
\Gamma(n \rightarrow \pi^0 e_i^+)
            &=& \frac{m_n}{32\pi f_{\pi}^2}
                \left|C(dude_i) \left[ 1+D+F \right] \right|^2\\
\Gamma(p \rightarrow \pi^0 e^+)
            &=& \frac{m_p}{64\pi f_{\pi}^2}
                \left(\left|\tilde{C}^{(1)}\right|^2 + \left|\tilde{C}^{(2)}
		\right|^2\right)
                \left[ 1+D+F \right]^2
\end{eqnarray*}
\normalsize

\newpage

\section*{Figure Captions}
\renewcommand{\labelenumi}{Fig.~\arabic{enumi}}
\begin{enumerate}
\item A supergraph contributing to the dimension-five operators of the nucleon
decay.

\item Allowed ranges on the color-triplet Higgs mass $M_{H_C}$ and the
``GUT-scale'' $M_{GUT} \equiv (M_V^2 M_\Sigma)^{1/3}$ obtained from the
renormalization group analysis (thick lines), by varying $m_{\tilde{h}}$ and
$m_{\tilde{g}}$ between 100~GeV and 1~TeV.  $M_{H_C}$ depends only on
$m_{\tilde{h}}$, and $M_{GUT}$ only on $m_{\tilde{g}}$. We use the gauge
coupling constants at the weak-scale given in the text. Also shown are the
ranges with an improved measurement on the strong coupling constant, $\alpha_3
= 0.118 \pm 0.0035$ (thin lines).

\item Lower bound on $M_{H_C}$ derived from the nucleon-decay experiments.
The horizontal axis represents $|1 + y^{tK}|$, the sum of the second- and
third-generation contributions normalized by the second-generation one. The
vertical axis corresponds to $M_{H_C}$. The shaded region is excluded. The
upper curve corresponds to the hadron matrix element $\beta = 0.03$~GeV$^3$,
the lower one to $\beta = 0.003$~GeV$^3$. The experimental limits come from the
mode $n \rightarrow K^0 \bar{\nu}_\mu$ for $|1 + y^{tK}| > 0.4$, and from the
mode $n \rightarrow \pi^0 \bar{\nu}_\mu$ for $|1 + y^{tK}| < 0.4$ and
$|1+y^{t\pi}|=1$.  The short-range renormalization factor $A_S$ is taken to be
$A_S=0.67$.  The maximum value on $M_{H_C}(=2.3\times 10^{17}~\GeV)$ from the
renormalization-group (RG) analysis requiring gauge coupling unification (see
section 4) is also shown.

\item The dependence of the lower bound of $M_{H_C}$ on the parameters
$m_{\tilde{Q}}$ and $m_{\tilde{w}}$. The dashed line shows the dependence on
$m_{\tilde{w}}$ taking $m_{\tilde{Q}}=1~\TeV$.  The dash-dotted line shows the
dependence on $m_{\tilde{Q}}$ when $m_{\tilde{w}}=45~\GeV$.  In both curves we
have taken the most conservative set of parameters, $\tan\beta_H=1$,
$|1+y^{tK}|<0.4$, $|1+y^{t\pi}|=1$, $A_S=0.67$, and $\beta=0.003~\GeV^3$. We
have assumed $m_{\tilde{L}}\simeq m_{\tilde{Q}}$. The maximum value on
$M_{H_C}(=2.3\times 10^{17}~\GeV)$ from the renormalization-group (RG) analysis
requiring gauge coupling unification (see section 4) is also shown.

\item The dependence of the lower bound of $M_{H_C}$ on $\tan\beta_H$.
The upper curve is obtained with $|1+y^{tK}|=1$, and the lower curve with
$|1+y^{tK}|<0.4$ and $|1+y^{t\pi}|=1$. We have taken
$m_{\tilde{Q}}=m_{\tilde{L}}=1~\TeV$, $m_{\tilde{w}}=45~\GeV$, $A_S=0.67$ and
$\beta=0.003 ~\GeV^3$. The maximum value on $M_{H_C}(=2.3\times 10^{17}~\GeV)$
from the renormalization-group (RG) analysis requiring gauge coupling
unification (see section 4) is also shown.

\item The limits on $m_{\tilde{w}}$ and $m_{\tilde{Q}}$
from the KAMIOKANDE nucleon-decay experiments, in the absence of the
cancellation between second- and third-generation contributions ({\it i.e.},\/
$|1+y^{tK}|=1$). The most conservative parameters, $M_{H_C}=2.3\times
10^{17}~\GeV$, $\tan \beta_H=1$, $\beta =0.003$~GeV$^3$, and $A_S=0.67$, are
used. We have assumed $m_{\tilde{Q}} \simeq m_{\tilde{L}}$ for simplicity. The
shaded region is excluded. Also shown are the limits from the direct search
experiments on wino and squarks at LEP and CDF.

\item The same as in Fig.~6, but allowing the cancellation between second- and
third-generation contributions ({\it i.e.}, $|1+y^{tK}|<0.4$,
$|1+y^{t\pi}|=1$).

\item The same as in Fig.~6, but with an improved constraint by a factor of 30
expected at superKAMIOKANDE. The expected limit on $m_{\tilde{w}}$ from the
LEP-II experiment ($m_{\tilde{w}} > 90~\GeV$) is also shown.

\item The same as in Fig.~7, but with an improved constraint by a factor of 30
expected at superKAMIOKANDE. The expected limit on $m_{\tilde{w}}$ from the
LEP-II experiment ($m_{\tilde{w}} > 90~\GeV$) is also shown.

\item The same as in Fig.~9. We have assumed that the error-bar of $\alpha_3$
is reduced by a factor of 2 with the same central value, leading to a stronger
upper bound on $M_{H_C} (< 6.1 \times 10^{16}~\GeV)$ (see Fig.~2).

\item The renormalization factor $A_S$ vs. $m_t/\sqrt2\sin\beta_H$. The solid
line represents the $A_S$ for the dimension-five operators only with first- and
second-generation fields, and dash-dotted line for the operator $(Q_t Q_t)
(Q_c L_\mu)$. The upper horizontal line is $A_S$ derived by the authors in
Ref.~\cite{ENR}.  The lower horizontal line is $A_S$ which does not contain the
contribution of the top-quark Yukawa coupling ({\it i.e.},\/ $m_t = 0$).

\end{enumerate}

\end{document}